%% file: main.tex
\documentclass[pdflatex,sn-nature]{sn-jnl}

\usepackage{xargs}%
\usepackage{svg}%
\usepackage{tabularx}
\usepackage{bm}%
\usepackage{graphicx}%
\usepackage{multirow}%
\usepackage{amsmath,amssymb,amsfonts}%
\usepackage{amsthm}%
\usepackage{mathrsfs}%
\usepackage[title]{appendix}%
\usepackage{xcolor}%
\usepackage{textcomp}%
\usepackage{manyfoot}%
\usepackage{booktabs}%
\usepackage[linesnumbered,ruled,vlined]{algorithm2e}

\SetCommentSty{mycommfont}

\SetKwInput{KwInput}{Input}                
\SetKwInput{KwOutput}{Output}              
\SetKwInput{KwParam}{Parameters}              
\SetKwRepeat{Do}{do}{while}%
\usepackage{listings}%
\usepackage{cleveref}%
\usepackage{adjustbox}%
\theoremstyle{thmstyleone}%
\theoremstyle{thmstyletwo}%

\theoremstyle{thmstylethree}%

\usepackage{tikz}
\usepackage{pgfplots}
\usepackage{adjustbox}
\usetikzlibrary{calc, positioning, arrows.meta, shapes.geometric}

\usepackage{soul}

\newcommand{\dA}{\textrm{A}}
\newcommand{\dC}{\textrm{C}}
\newcommand{\dG}{\textrm{G}}
\newcommand{\dT}{\textrm{T}}

\newcommand{\bone}{\textcolor{blue}{1}}
\newcommand{\bzero}{\textcolor{blue}{0}}

\crefname{figure}{Figure}{Figures}
\usepackage[colorinlistoftodos,prependcaption,textsize=tiny]{todonotes}

\input{resources/macros}

\begin{document}

\title[Article Title]{SynDe: Syndrome--guided Decoding of Raw Nanopore Reads}


\author*[1]{\fnm{Anisha} \sur{Banerjee}}\email{anisha.banerjee@tum.de}

\author[2,3]{\fnm{Roman} \sur{Sokolovskii}}

\author[2]{\fnm{Thomas} \sur{Heinis}}%
\author[1]{\fnm{Antonia} \sur{Wachter-Zeh}}%
\author[4]{\fnm{Eirik} \sur{Rosnes}}%
\author[5]{\fnm{Alexandre} \sur{Graell i Amat}}

\affil[1]{ Institute for Communications Engineering; Technical University of Munich (TUM), Munich, Germany
}

\affil[2]{Department of Computing, Imperial College London, UK
}

\affil[3]{European Molecular Biology Laboratory, European Bioinformatics Institute (EMBL-EBI), Wellcome Genome Campus, Hinxton, UK
}

\affil[4]{Simula UiB, Bergen, Norway
}

\affil[5]{Department of Electrical Engineering, Chalmers University of Technology, Gothenburg, Sweden
}


\abstract{ 
    Nanopore sequencing technology remains highly error-prone, making efficient error correction essential in DNA-based data storage. Prior work addressed high error rates  using convolutional codes with their decoder coupled with the basecaller
    , but such approaches only accommodate a limited number of code classes and incur significant decoding complexity. To overcome these limitations, we propose two algorithms: \ps, which efficiently detects primer sequences in raw nanopore sequencing reads, and \syndec, a decoder that operates on the same raw reads and supports any linear error correction code with a low-complexity graphical representation.  
    
     \ps~provides primer location estimates close to those of existing approaches while being better suited for real-time primer detection during sequencing. \syndec~performs well with convolutional codes augmented with periodic markers, often approaching or exceeding the performance of existing algorithms with a lower time complexity. Remarkably, the confidence scores produced by \syndec~reliably identify which of its outputs should be discarded.  
    
}

\keywords{DNA data storage, Nanopore sequencing, Convolutional codes, basecaller-decoder integration}

\maketitle

\section{Introduction}\label{sec:intro}



The promise of DNA-based data storage is unprecedented density, durability, and low storage maintenance costs.
However, despite the existence of high-performance end-to-end solutions~\cite{lauMagneticDNARandom2023, welterEndtoEndCodingScheme2024, welzelDNAAeonProvidesFlexible2023}, the high costs of DNA synthesis and sequencing with current technology remain a significant barrier to its widespread adoption.
At the reading end, robust information recovery typically requires sequencing several copies of the same synthesized DNA sequence (a parameter referred to as coverage), which both decreases the physical information density and increases the computational cost of decoding.
This problem is especially relevant in the case of relatively error-prone nanopore sequencing~\cite{deamerThreeDecadesNanopore2016, mackenzieIntroductionNanoporeSequencing2023, maoModelsInformationTheoreticBounds2018}, which has gained increased prominence due to its falling costs, the ability to perform on-the-fly data analysis and decoding, and the potential to support a wide range of throughput targets from portable solutions~\cite{yazdi_portable_2017} to large-scale deployments. However, it suffers from higher basecalling error rates than Illumina or PacBio sequencing~\cite{delahayeSequencingDNANanopores2021}.

A promising approach to addressing the high error rates and computational cost of nanopore sequencing in DNA storage pipelines is to integrate error correction decoding directly into the basecalling process, leveraging the rich features of raw reads from the nanopore device.
(Or, as in~\cite{vidalConcatenatedNanoporeDNA2024}, bypassing basecalling altogether.)
The pioneering work by Chandak \textit{et al.}~\cite{chandakOvercomingHighNanopore2020} presented a decoding framework, referred to as \chandakdec~in \cite{volkel_nanopore_2025} and in this work, that integrates list decoding of convolutional codes into the Flappie basecaller (an extension to the Bonito basecaller has also been provided in \cite{lauMagneticDNARandom2023}). 
Volkel \textit{et al.}~\cite{volkel_nanopore_2025} continued this line of research by proposing \volkeldec, which adapts the convolutional decoding of HEDGES~\cite{press2020hedges}---a constrained\footnote{Relatedly, the integration of biologically-motivated constraints, such as homopolymer runs or GC-balance, into basecalling has been investigated in~\cite{menon2025basecalling}.} convolutional code---into the Bonito basecaller. 

The error correction capability of both frameworks is determined by the \emph{memory} of their underlying convolutional codes, i.e., the number of memory elements (delay units) in the encoder,  which determines how many past message symbols influence each codeword  symbol. Typically, a code with higher memory is more resilient to errors, but at the cost of a higher decoding complexity. This trade-off exposes a key limitation of the works~\cite{chandakOvercomingHighNanopore2020,volkel_nanopore_2025}: both  decoders operate on the full graphical representation of the convolutional codes to perform an  exhaustive search for the most likely codeword. Consequently, their computational complexity grows exponentially with the memory. 
The second problem is related to the fact that a basecaller-integrated decoder needs to know the starting position of the codeword in the raw read.
To that end, both approaches basecall the entire read first (without decoder integration) and then attempt to identify predetermined flanking sequences that are incorporated in the synthesized strand.
(These sequences are often used for amplification via polymerase chain reaction for random access~\cite{organick_random_2018}; therefore, here we refer to them as primers.)
This means that in their pipelines, basecalling has to be performed twice: once on the full read to find the primers, and the second time, in conjunction with error correction decoding, to recover the stored data.
Besides incurring an additional computational cost, this also prevents these decoding schemes from potentially being used on-the-fly during the sequencing run itself.

In this work, we advance research on basecaller-decoder integration by directly addressing the two challenges above. First, we propose a sequential 
decoding scheme, named \syndec, that relies on the \textit{syndrome trellis} representation of convolutional codes~\cite{wolfEfficientMaximumLikelihood1978, bahlOptimalDecodingLinear1974, sidorenkoDecodingConvolutionalCodes1994, rosnesMaximumLengthConvolutional2004} and incorporates it into the beam search algorithm, which underpins most basecalling pipelines~\cite{scheidlWordBeamSearch2018, gravesConnectionistTemporalClassification2006, freitagBeamSearchStrategies2017}. 
The main advantage of the proposed scheme is that its complexity is independent of the memory 
of the convolutional code and linear in its length; in Section~\ref{sec:results}, we show how the use of higher-memory codes allows us to approach the frame error rate (FER) performance of \chandakdec~with lower complexity. In addition to lower complexity, our decoding scheme is more universal than the existing alternatives, in the sense that unlike \chandakdec~and \volkeldec, it can flexibly incorporate any linear error correction code with a compact syndrome trellis and is easily adapted to the inclusion of markers or other coding-theoretic techniques, such as puncturing, without bespoke modification of the decoding scheme for each setup.

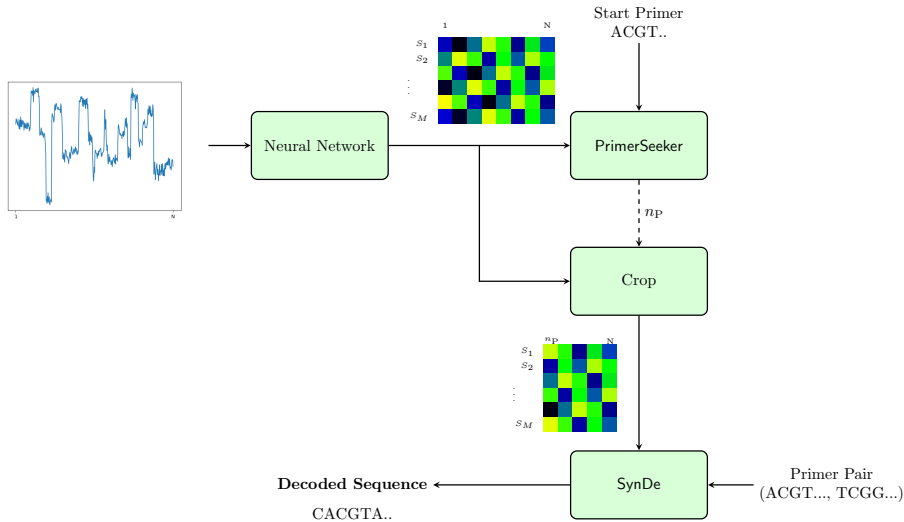
\begin{figure}
    \centering
    \scalebox{0.6}{\input{resources/decoding-pipeline/pipeline}}
    \caption{Overview of our decoding workflow. A raw signal is first processed by a neural network that generates a probability matrix over all possible states (for instance $k$-mers or CTC tokens). This matrix is fed to our \ps~algorithm, which estimates the position at which the primer sequence starts in the raw read, say $n_{\textnormal{P}}$. The columns preceding the $n_{\textnormal{P}}$-th column are  cropped and the trimmed probability matrix is passed to \syndec, which leverages the syndrome trellis of the adopted convolutional codes and incorporates knowledge of any included marker symbols to perform ``code-aware'' basecalling, ultimately generating the decoded DNA sequence. 
    }
    \label{fig:pipeline}
\end{figure}

Second, we propose a dedicated algorithm that locates the start of the flanking sequences in the raw read. We refer to this as \ps, and unlike alternative solutions, it does not require basecalling the entire raw signal. Its computational complexity is nearly identical to that of full basecalling, thus making it very well-suited for online decoding workflows. In essence, \ps~is a modified beam search where the beams are restricted to the targeted primer sequence and are well-separated in their starting positions (see Section~\ref{sec:results}). Our complete decoding workflow is shown in Fig.~\ref{fig:pipeline}.

Taken together, these contributions pave the way for computationally efficient integration of decoding and basecalling in future DNA storage systems.

\section{Results} \label{sec:results}



\subsection{Efficient Localization of Primer Sequences in Raw Reads} \label{sse:results_primerseeker}

Existing approaches to basecaller-decoder integration rely on full basecalling for identifying the beginning of the codeword. This is a computational bottleneck that prevents on-the-fly decoding. Here, we develop \ps, an algorithm that efficiently detects and localizes a given target sequence directly in raw nanopore signals (Fig.~\ref{fig:primer_search}, Methods \ref{methods::primer-search}).

\begin{figure}
    \centering
  \begin{tikzpicture}
      \begin{axis}[legend style={nodes={scale=0.8, transform                shape}},
				xmin=0, xmax=60,
				xlabel=\textsc{$\Delta$}, ylabel={Fraction of reads with distance exceeding $\Delta$},
				legend pos=south east, legend style={font=\scriptsize},
				grid=both]
            \addplot[solid, blue!60!black, mark=diamond*, mark options={fill=blue!60!black}, restrict x to domain=0:60] table {resources/primer-search-plots/primerseeker_volkel_ctc_vs_chandak.txt};
            \addlegendentry{A: \psctc~v/s Chandak \textit{et al.} \cite{chandakOvercomingHighNanopore2020}};
            \addplot[solid, red!80!black, mark=square*, mark options={fill=red!80!black}, restrict x to domain=0:60] table {resources/primer-search-plots/primerseeker_volkel_lokatt_vs_guppy.txt};
            \addlegendentry{B: \pslokatt~v/s Guppy};
        \end{axis}
  \end{tikzpicture}
    \caption{Performance of the primer search algorithm on the dataset from Volkel \textit{et al.} \cite{volkelNominalFAST5FASTQ2024}. In both curves A and B, two methods of locating a target sequence in raw reads are compared. For each positive integer $\Delta$ (x-axis), the y-axis shows the fraction of raw reads for which the position estimates from the competing methods lie within $\Delta$ samples of each other. \textbf{A}: \psctc~when compared against estimates obtained by the method used by \chandakdec. \textbf{B}: Comparison between \pslokatt~and estimates obtained using the Guppy basecaller. The Guppy basecaller generates an erroneous sequence along with the raw signal positions of the corresponding base transitions. The starting position of the primer is considered to be the raw read sample that corresponds to the transition into the first base of the basecalled sequence, which matches the primer sequence most closely. 
    }
    \label{fig:primer-search-sims}
\end{figure}
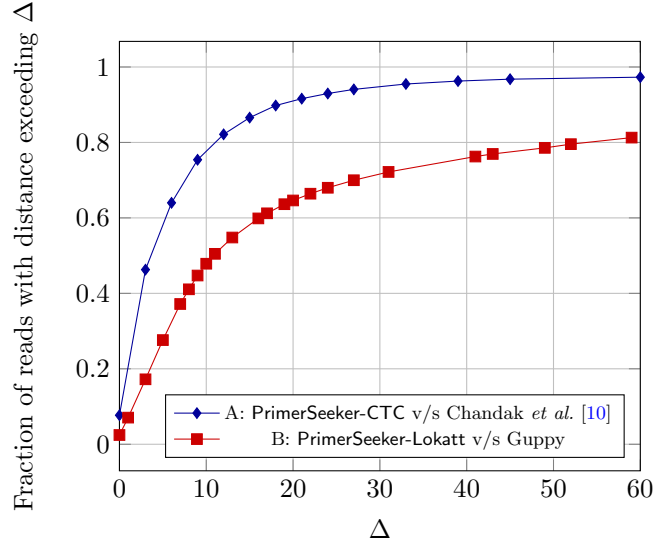

On the R9.4.1 chemistry dataset by Volkel \textit{et al.}~\cite{volkel_nanopore_2025}, \pslokatt, an implementation of \ps~based on the Lokatt basecaller~\cite{xuLokattHybridDNA2023} (Methods~\ref{sec:basecalling-beam-search}), achieves $78.9\%$ agreement within $50$ samples ($\sim\,$$5$ nucleotides) with the standard approach based on full Guppy basecalling~\cite{GuppyProtocol2018} (Fig.~\ref{fig:primer-search-sims}A). 
We also evaluate a variant of \ps~based on the Bonito basecaller \cite{oxfordnanoporetechnologiesBonito2025}, referred to as \psctc. 
This method demonstrates $97\%$ agreement within $50$ samples ($\sim\!5$ nucleotides) when compared with the more stringent ``localization-by-basecalling'' approach in~\cite{chandakOvercomingHighNanopore2020, lauMagneticDNARandom2023}, which attempts to locate \textit{both} flanking sequences (Fig.~\ref{fig:primer-search-sims}B).\footnote{We remark that the probability matrices generated by the CTC-based neural network only account for a part of the raw read. Therefore, it is not possible to compare \psctc~with \pslokatt~directly.}  This performance is achieved with a computational complexity nearly identical to that of basecalling (Supplementary~\ref{supp:complexity-primer-search}). Moreover, unlike conventional approaches that  require basecalling the complete signal, \psctc~enables reliable online detection of target sequences. 

\begin{figure}
    \centering
    \scalebox{0.5}{\input{resources/primer-search-pipeline/primer_search_alg}}
    \caption{Illustration of \pslokatt. This example considers $\dC\dA\dC\dG\dT\dA\dG\dG$ as the primer and $k=2$. A neural network takes the raw signal as input and returns a probability matrix, say $\boldsymbol{P}$, where the $(i,j)$-th entry denotes the probability that the $i$-th sample resulted from the presence of the $j$-th $k$-mer in the pore, for all possible $k$-mers. This matrix is used by the primer search algorithm, which, for instance, considers two candidate starting positions in the raw read, $a$ and $b$. Starting with two initial beams, one for each candidate, the algorithm extends each by the first $k$-mer $\dC\dA$ and scores them according to $\boldsymbol{P}$. Each beam is subsequently propagated through the upcoming samples through either (i) a dwell event, where the current $k$-mer remains in the pore for an additional sample; or (ii) an extension event, where the next nucleotide translocates into the pore, thus transitioning to the next $k$-mer state. At each iteration, the algorithm propagates all beams into two child beams, which are assigned likelihood scores using $\boldsymbol{P}$, and merging any beams that share identical starting and ending positions and represent the same sequence of $k$-mers. Whenever a beam has traversed all $k$-mers corresponding to the target primer sequence, its final score is considered to be the probability that the primer sequence began at its starting position. The algorithm outputs the start position associated with the beam having the highest score. 
    }
    \label{fig:primer_search}
\end{figure}

\subsection{Low-complexity Integration of Basecalling and Decoding}
\label{sec::integrated-decoding}




The main basecaller-decoder integration approaches in the literature, namely, 
\chandakdec~\cite{chandakOvercomingHighNanopore2020, lauMagneticDNARandom2023} and \textsf{AlignmentMatrix} \cite{volkel_nanopore_2025}, achieve high decoding accuracy but their complexity scales exponentially with the memory (Supplementary~\ref{supp:complexity-synde}), limiting their practical application. Here, we propose \syndec~(Methods~\ref{methods::decoding}), a decoding algorithm that relies on the \textit{syndrome trellis} of a linear code (Methods~\ref{sec:conv-codes}). \syndec's time complexity is linear in the length of the raw read and independent of the memory of the code (Supplementary~\ref{supp:complexity-synde}).

\begin{figure} 
    \centering
    \scalebox{1.2}{\begin{tikzpicture}
      \begin{axis}[legend style={nodes={scale=0.55, transform shape}},
				ymode=log,
				xmin=-7, xmax=105,
				ymin=1.001e-3, ymax=1.0,
				xlabel={\small Percentage of reads discarded (\%)}, ylabel={\small Frame Error Rate},
				legend pos=south west, legend style={font=\scriptsize},
				grid=both]
            \addplot[dashed, green!40!black, mark=triangle*, mark options={fill=green!40!black}] table {resources/fer-discarded-reads-plots/volkel_dataset/FER_Stan_cc4_3_8_r0.68.txt};
            \addlegendentry{\chandakdec, CC8-3, rate=$0.678$};
            \addplot[dash dot, red!90!black, mark=*, mark options={fill=red!90!black}] table {resources/fer-discarded-reads-plots/volkel_dataset/FER_CTC_cc13_12_10_m6_r0.744.txt};
            \addlegendentry{\syndecctc, CCM10-6, rate=$0.744$};
            \addplot[dash dot, blue!90!black, mark=*, mark options={fill=blue!90!black}] table {resources/fer-discarded-reads-plots/volkel_dataset/FER_CTC_cc9_8_9_m5_r0.702.txt};
            \addlegendentry{\syndecctc, CCM9-5, rate=$0.702$};
        \end{axis}
  \end{tikzpicture}}
    \caption{Comparison of the decoding performance of \syndecctc~with \chandakdec~\cite{chandakOvercomingHighNanopore2020} on the dataset from Volkel \textit{et al.} \cite{volkelNominalFAST5FASTQ2024}. The x-axis values indicate the percentage of raw reads that were discarded, while the corresponding y-axis values indicate the FER, i.e., the fraction of the remaining reads that were decoded incorrectly. We observe that for the same fraction of discarded reads, \syndecctc~achieves a better FER than \chandakdec~despite using codes of slightly higher rates. The identifier of each code CC(M)x-y suggests that the associated memory is x. More details on the codes can be found in Supplementary~\ref{supp:encoding} (Table~\ref{tab:codes1} and Table~\ref{tab:codes2}).  }\label{fig:fer-vs-discarded-reads}
\end{figure}
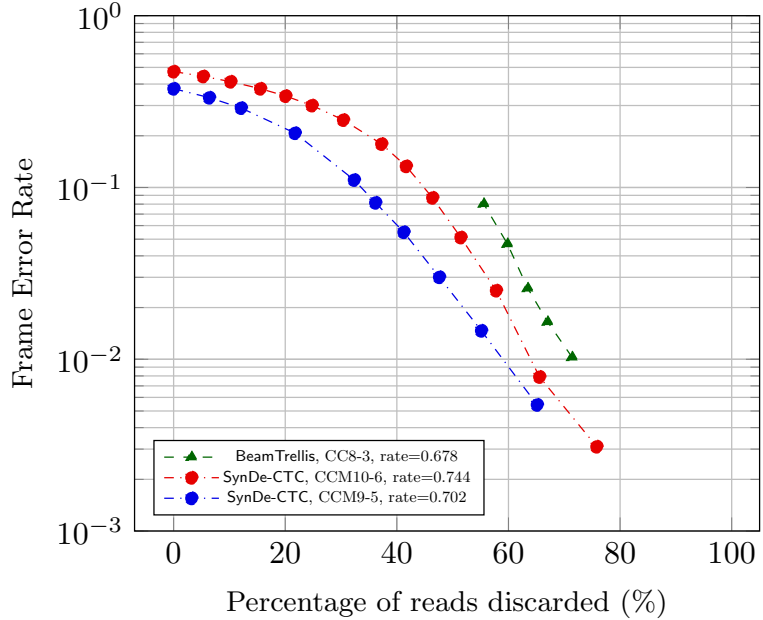

Decoding via the syndrome trellis allows us to use error correction codes with larger memory than in existing schemes and allows a greater flexibility in the choice of the code family. We opt for a combination of convolutional and marker codes (Methods~\ref{sec:conv-codes})---marker codes provide a synchronisation mechanism against insertion/deletion errors, whereas convolutional codes correct substitutions. On the R9.4.1 dataset by Volkel \textit{et al.}~\cite{volkel_nanopore_2025}, \syndecctc~(our Bonito-based implementation of \syndec) outperforms \chandakdec~across multiple configurations (Fig.~\ref{fig:fer-vs-discarded-reads}). For instance, at $65\%$ reads discarded and code rate $\sim\!0.7$, \syndecctc~achieves a FER of $0.5\%$  against \chandakdec's $1.5\%$ FER at code rate $\sim\!0.68$---a $3$-fold improvement in accuracy at lower redundancy cost and considerably lower computational complexity. Further results can be found in Supplementary~\ref{supp::sec::synde-perf}.

\section{Discussion} \label{sec:discussion}

This work proposes an on-the-fly 
decoding pipeline that operates directly on raw reads from nanopore sequencers. As depicted in Fig.~\ref{fig:pipeline}, this pipeline consists of two major components. The first component, named \ps, focuses on efficiently detecting and locating target sequences within the raw reads, while the second, referred to as \syndec, achieves basecaller-decoder integration with a sequential algorithm based on beam search. 

Detection of target sequences in raw reads is a critical task in DNA-based data storage. However, to the best of our knowledge, no prior work has attempted this detection directly in the raw signal---i.e., without resorting to full basecalling prior to detection. 
The current state-of-the-art relies on full basecalling, followed by searching the basecalled sequence for regions that are similar to the target sequence, typically using metrics such as edit distance. This approach incurs the computational burden of both basecalling and subsequent sequence alignment or edit distance computations. We demonstrate on the dataset by Lau \textit{et al.} that an optimized version of \psctc~achieves a computational complexity nearly identical to that of basecalling (see Supplementary~\ref{supp:complexity-primer-search}) while closely matching the results of the primer localization technique employed by Chandak \textit{et al.}\footnote{On Lau \textit{et al.}'s dataset, which contains reads of short DNA strands ($\sim\!150$ nts), this comparison involves both approaches seeking the length-$23$ suffix of the $25$-nts primer. The first $2$ nucleotides are ignored to prevent any false detections arising from boundary effects or potential cropping in the nanopore raw signal.}, despite \psctc~not exploiting the primer sequence at the other end as an anchor, or the knowledge of the length of the payload. This suggests that \ps~is uniquely well-suited for use in online readout pipelines for DNA data storage.


Once the starting position of the primer sequence is detected, \syndec~focuses on the part of the raw read associated with the payload and its surrounding primers, and starts performing a constrained form of basecalling, which exclusively looks for codewords instead of all possible DNA sequences. It does so by conducting a sequential exploration of the syndrome tree of the encoder, in a manner similar to that of basecalling, implying that the decoding complexity of \syndec~is independent of the encoder's memory (Supplementary~\ref{supp:complexity-synde}). In other words, the computational complexity of \syndec~only scales linearly with the length of the payload. As a result, \syndec~is able to recover the information content of the payload at significant complexity gains over \chandakdec~and \volkeldec, both of which have complexities that scale exponentially with memory (see Supplementary~\ref{supp:complexity-synde}). Therefore, \syndec~is the first solution for basecaller-decoder integration that enables the use of high-memory convolutional codes, which are known to offer improved error-correction performance. 


Another drawback of \chandakdec~is that it is restricted to convolutional codes with parameters $(2,1)$, i.e.,  encoders that produce two codeword symbols per message symbol. To obtain convolutional codes of higher rates, \chandakdec~applies  \emph{puncturing} to a  mother $(2,1)$ convolutional code, i.e., it periodically deletes  selected output symbols. This is not ideal, as high-rate convolutional codes obtained via puncturing are generally known to exhibit inferior error-correction performance compared to native high-rate convolutional codes.
 \syndec~addresses this limitation by providing a versatile decoding framework  applicable to any linear error correction code admitting an efficient syndrome trellis representation, including  native higher-rate convolutional codes\footnote{To ensure that the storage complexity of the syndrome trellis does not exceed the capacity of the cache memory (say $32$MB), our implementation of \syndec~can only simulate a convolutional code of memory at most $18$. Similar arguments lead us to conclude that any linear code with at most $120$ quaternary symbols and rate at least $0.93$, can be used with \syndec.}(Supplementary~\ref{supp:decoding-sims})  as well as more general linear codes. Thus, \syndec~is capable of accommodating a much broader class of linear codes than \chandakdec. Furthermore, by additionally incorporating marker symbols that help synchronize with the raw read, \syndec~is usually able to avoid errors arising from dwell time mispredictions (Supplementary~\ref{supp:decoding-sims}). 
Indeed, this flexibility enables us to easily and cleverly tune the trade-off between redundancy allocated to synchronization (via marker symbols) and redundancy allocated to correcting substitution errors (via the convolutional code).

It is also worth mentioning that although we left the processing of reverse-complemented reads outside the scope of the paper, they can be straightforwardly accommodated using the same technique as in \cite{chandakOvercomingHighNanopore2020, lauMagneticDNARandom2023}.


An interesting direction for future work is to incorporate into \syndec~ideas from Fano's decoding algorithm \cite{fanoHeuristicDiscussionProbabilistic1963}, which operates on a similar graphical representation of convolutional codes but employs a more sophisticated (and costly) beam scoring method: it not only considers the likelihood of past observations matching the beam's codeword symbols, but also accounts for the probability of \textit{future} observations aligning with a valid codeword. The task of tailoring popular iterative decoding schemes from coding theory to exploit the soft information in raw reads, as done here, is also an open problem that deserves attention.

Overall, \syndec~is the first basecaller-decoder integration algorithm with time complexity scaling linearly in the code length and comparable-to-superior performance relative to existing approaches. In conjunction with \ps, the two algorithms developed here provide an efficient readout pipeline for DNA-based data storage.


\section{Methods}
\subsection{Convolutional \& Marker Codes} \label{sec:conv-codes}

Our error correction scheme uses convolutional codes, a class of linear codes with encoders that involve memory. 
Specifically, the encoder of a $(c, b, \nu)$ convolutional code can be implemented as a circuit of $\nu$ shift registers, which for every $b$ message symbols, produces $c > b$ encoded symbols. Each block of $c$ output symbols depends not only on the current $b$ inputs, but also the past $\nu$ message symbols. For this reason, $\nu$ is referred to as the \emph{memory} of the code. 

Convolutional codes are widely used due to their structural properties, which enables efficient maximum-likelihood decoding. In this work, we focus on the syndrome trellis representation of these codes, which exploits the fact that every codeword of a linear code $\cC$ is orthogonal to every codeword of the dual code, the basis of which is known.

More specifically, each codeword $\bfc$ of a convolutional code satisfies the equation $\bfc\boldsymbol{H}^T = \boldsymbol{0}$, where the rows of the matrix $\boldsymbol{H}$, also known as the \emph{parity-check matrix}, specify the basis of the dual code. This phenomenon allows us to represent all codewords as paths through a \emph{syndrome trellis}. Such a graph comprises multiple nodes, each representing a specific \emph{syndrome state}, distributed across $N+1$ levels, where $N$ is the code length. In particular, the codeword $\bfc$ traces a path of syndrome states, starting with $s_0(\bfc)=\bfzero$ at the $0$-th level and subsequently passing through the states $s_t(\bfc) = c_1\bfh_1+\cdots+ c_t\bfh_t$ for the levels $t=1,\ldots,N$, where $\bfh_i$ denotes the $i$-th column of $\boldsymbol{H}$. By definition, the path terminates at the syndrome state $s_N(\bfc) = \bfzero$. 

To exemplify these concepts, we consider a binary convolutional code from \cite{rosnesMaximumLengthConvolutional2004} of length $N=10$ with $c=4$, $b=2$,  $\nu=3$, and the  parity-check matrix 
\begin{equation}
    \boldsymbol{H} = \begin{bmatrix}
        1 & 1 & 0 & 0 & & & & & & \\
        0 & 1 & 1 & 1 & & & & & &  \\
        1 & 1 & 1 & 1 & \bone & \bone & \bzero & \bzero & &  \\
        0 & 1 & 0 & 1 & \bzero & \bone & \bone & \bone & &  \\
        0 & 0 & 0 & 0 & \bone & \bone & \bone & \bone & &  \\
        0 & 0 & 1 & 1 & \bzero & \bone & \bzero & \bone & &  \\
         &  &  &  & \bzero & \bzero & \bzero & \bzero & 1 & 1 \\
         &  &  &  & \bzero & \bzero & \bone & \bone & 0 & 1
    \end{bmatrix}. \label{eq::parity}
\end{equation}

As is common with convolutional codes, we observe that $\boldsymbol{H}$ is composed of repeated instances of a smaller submatrix (indicated in blue in (\ref{eq::parity}))
, which implies that for any integer $t$ ranging from $1$ to $N$, $s_t(\bfc)$ is a binary vector that typically has at most $\nu + c-b = 5$ non-zero entries. Consequently, the memory of a code directly influences the complexity of the syndrome trellis, i.e., the number of possible realizations of the syndrome state $s_t(\bfc)$ for any $t$ in the range from $1$ to $N$.



\begin{figure}
    \centering
    \scalebox{0.8}{\input{resources/decoding-pipeline/syndrome-trellis}}
    \caption{Syndrome trellis of a binary convolutional code of length $N=10$ with parity-check matrix given by Eq.~(\ref{eq::parity}). Every codeword of this code, say $\bfc$, corresponds to a specific path from the starting node to the terminating node. If the $i$-th edge of the path traced by $\bfc$ is dashed, $c_i=0$, and $c_i=1$ otherwise. Each node represents a specific syndrome state, the (compressed) decimal representation of which indicated by the labels of the nodes. For instance, the path marked in green corresponds to the codeword $\bfc=0011111100$. Note that we draw the binary trellis here for illustrative purposes; the quaternary trellis can be straightforwardly obtained by merging every two consecutive trellis sections. In such a quaternary trellis, we may apply the mapping $00\xrightarrow[]{} \dA$, $11\xrightarrow[]{} \dT$ to transform the codeword $\bfc$ into the DNA codeword $\dA\dT\dT\dT\dA$.
    }
    \label{fig:syn-trellis}
\end{figure}
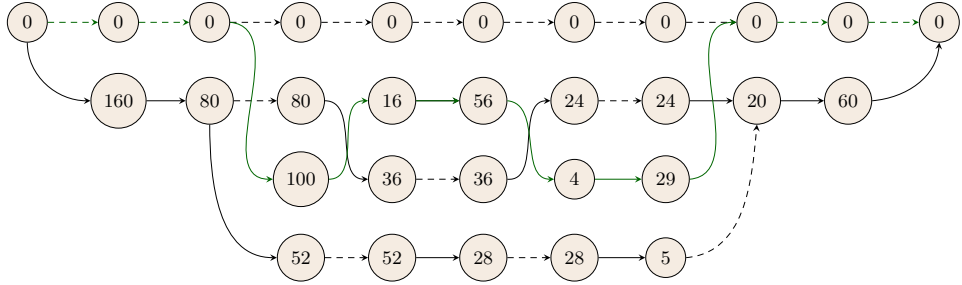

The syndrome trellis corresponding to our example code is illustrated in Fig.~\ref{fig:syn-trellis}. Each node in the trellis represents a specific syndrome state $s_t(\boldsymbol{c}) \in \{0,1\}^8$, the decimal representation of which is indicated in the node labels. A convenient feature of such binary syndrome trellises is that they can be effortlessly transformed into a syndrome trellis over the quaternary alphabet, as necessitated by DNA sequences, by simply merging two consecutive trellis sections into one. For instance, if we assume the mapping $00\xrightarrow[]{} \dA$, $11\xrightarrow[]{} \dT$,  and so on, then the green path through the syndrome trellis in Fig.~\ref{fig:syn-trellis} corresponds to the DNA sequence $\dA\dT\dT\dT\dA$, that traces a path through the nodes $0$, $0$, $16$, $4$, $0$, and $0$ in the quaternary syndrome trellis. 

Indeed, for the purpose of adapting beam search algorithms to exclusively look for codewords, we leverage the tree representation of a quaternary syndrome trellis. This approach of achieving basecaller-decoder integration clearly offers greater flexibility, since it can be used with a broader class of codes---the codes with compact syndrome trellises. 
In contrast, the scheme proposed by Chandak \textit{et al.} only permits the use of a narrow subclass of punctured convolutional codes, which potentially excludes codes that are better suited to the nanopore sequencing channel, as discussed in Section~\ref{sec:discussion}. 

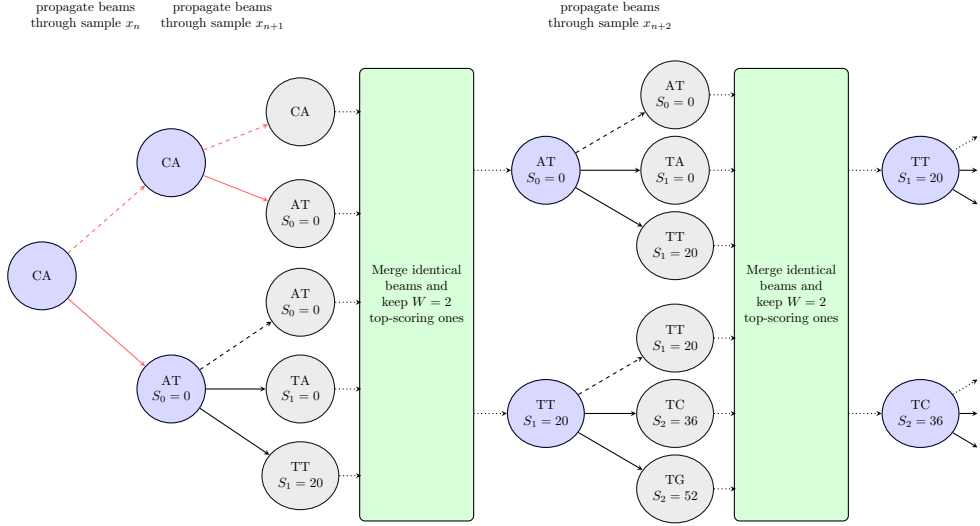
\begin{figure}
    \centering
    \input{resources/decoding-pipeline/beam-search-decoder}
    \caption{Illustration of \syndeclokatt~on a raw read $\boldsymbol{x}=(x_1,x_2,\ldots)$ for the convolutional code described in Fig.~\ref{fig:syn-trellis}. We assume $k=2$ and the primer sequence $\boldsymbol{p}=\dC\dA\dT$, which is estimated to start from the $n$-th sample of $\boldsymbol{x}$. \syndeclokatt~thus begins with a single beam corresponding to $\dC\dA$, the initial $k$-mer of $\boldsymbol{p}$, and propagates it through the upcoming samples in $\boldsymbol{x}$. As in Fig.~\ref{fig:primer_search}, every beam is propagated either by dwelling or by transitioning to the next permitted nucleotide, indicated by dashed and solid edges,  respectively. So long as a beam has not covered the complete $k$-mer sequence of $\boldsymbol{p}$, which is $(\dC\dA,\dA\dT)$, the beam is said to be in the `primer region' and the associated event edges are marked in red. Upon exiting the primer region, the beam enters the `payload region', where it is guided through the most promising codeword sequences by only examining those paths that are permitted by the quaternary syndrome trellis (the binary trellis is presented in Fig.~\ref{fig:syn-trellis}). In the payload region, each beam tracks the syndrome states traced by its hypothesized codeword in the quaternary syndrome trellis, indicated by the random variables $S_0,\ldots,S_i$, where $i$ refers to the number of codeword symbols covered by the beam so far. 
    }
    \label{fig:beam-search-decoder}
\end{figure}


Using convolutional codes alone for correcting synchronization errors, especially in conjunction with sequential algorithms like beam search, is generally a precarious endeavor
, for the following reason. Assume $k=2$ and that the true codeword contained in a raw read is given by $\dA\dT\dT\dT\dA$, prefaced by the primer sequence $\dC\dG\dA\dA$. Then \syndeclokatt~would propagate beams through the raw read samples as in Fig.~\ref{fig:beam-search-decoder}, hoping to have at least one beam that traverses the $k$-mer sequence $\dC\dG,\dG\dA,\dA\dA,\dA\dA,\dA\dT,\ldots$, 
that passes through the states $S_0=0,S_1=0,S_2=16,\ldots$ in the quaternary syndrome trellis. It might just happen that during this search, the beam corresponding to the (incomplete) $k$-mer sequence $\dC\dG,\dG\dA,\dA\dA,\dA\dA,\dA\dA$ (through the syndrome states $S_0=0,S_2=0,S_3=0$) 
survives the beam pruning step, while that corresponding to the sequence $\dC\dG,\dG\dA,\dA\dA,\dA\dA,\dA\dT$ does not due to inaccurate dwell time predictions. In this unfortunate event, our decoder has essentially fallen out of step with the syndrome trellis and this misalignment is catastrophic since the quaternary syndrome trellis will force this mis-synchronized beam to the codeword sequence $\dA\dA\dA\dA\dA$, even if the input probability matrix (which is agnostic to the convolutional code) might suggest that the DNA sequence $\dA\dA\dT\dT\dT\dA$ is more likely than the codeword $\dA\dA\dA\dA\dA$. Such errors are likely to be corrected by the standard decoding pipelines, where the decoder of a code works on the basecalled sequence.

To mitigate the possibility of falling out of step with the syndrome trellis, 
we employ marker codes \cite{ratzerMarkerCodesChannels2005}, which offer a simple and elegant solution to correct synchronization errors. Essentially, this involves inserting a short marker sequence in the codeword of a convolutional code after every $4$--$7$ symbols. This change is reflected in the syndrome trellis of the convolutional code by inserting `constant' edges corresponding to the marker sequence after each interval. Conducting the search over this modified trellis assists in discarding those beams, with high likelihood, that have lost synchronization with the syndrome trellis. 

Note that the presence of homopolymers in the codeword sequences also exacerbates the risk of misalignment of beams with the syndrome trellis. Thus, it is advisable to always apply a pseudo-random offset \cite{buttigiegCodebookMarkerSequence2011, daveyReliableCommunicationChannels2001} to a codeword sequence before generating the payload sequence for the DNA strands to be synthesized.

It is also worth noting that unlike Chandak \textit{et al.}, we do not append an $8$-bit cyclic redundancy check (CRC) to the message vector prior to convolutional encoding. While the CRC provides an integrity check by discarding codewords that fail validation, it naturally incurs a rate loss. In contrast, we exploit the likelihood scores provided by \syndec~and check if they lie above a specific threshold. In this manner, we avoid the redundancy overhead incurred by CRCs. Across all encoding schemes, the FERs of \syndecctc~decline as the threshold imposed on the output likelihood scores is increased; we therefore conclude that the decoder’s likelihood score serves as an effective measure of the reliability 
of the decoder output. 

The specific code constructions that we assess in this work are listed in Table~\ref{tab:codes1} and Table~\ref{tab:codes2}.

\subsection{Basecalling and Beam Search} \label{sec:basecalling-beam-search}

Since \syndec~is formulated by repurposing existing basecalling pipelines, we recall the core principles of the two basecallers which we focus on in this work.

\subsubsection{Lokatt Basecaller}

The Lokatt basecaller, recently proposed by Xu \textit{et al.} \cite{xuLokattHybridDNA2023}, models the process of nanopore sequencing as an explicit duration hidden Markov model (EDHMM). It assumes that the nanopore holds $k>1$ nucleotides within itself at any given time and that the duration for which a particular $k$-mer dwells in the pore follows some statistical distribution. Thus, the raw signal generated by the nanopore device can be seen as the output of an HMM that transitions between successive $k$-mers of the translocating DNA strand, with the added caveat that every state persists for a random amount of time. A given raw read, say $\boldsymbol{x}=(x_1,\ldots,x_N)$
, after some preprocessing, is fed to a neural network that estimates, for each current sample $x_i$, the probability that it resulted from the presence of a specific $k$-mer in the pore, for all possible $k$-mers. Thus, the network outputs a two-dimensional probability matrix, say $\boldsymbol{P}$, of $4^k$ rows and $N$ columns, that is used by the greedy marginalized beam search (GMBS) algorithm \cite{xuMarginalizedBeamSearch2024} to infer the most likely sequence of nucleotides that traversed the nanopore. Similar to the Viterbi and the beam search algorithms used by most basecallers, the GMBS algorithm estimates the most likely sequence of nucleotides by searching for the sequence of $k$-mers, say $\boldsymbol{k}$, that maximizes 
an approximation to 
the conditional probability $p(\boldsymbol{k}|\boldsymbol{x})$ under the EDHMM.

\subsubsection{CTC Basecallers}

Most basecallers for nanopore sequencing used today employ Connectionist Temporal Classification (CTC), a framework originally developed for sequence-to-sequence problems such as speech recognition. This technique obviates the need for segmenting the raw signal into the individual relatively stationary events, that correspond to the dwelling period of each $k$-mer. 

Under this setting, the raw signal is imagined to be the output of a state transition process, where the states are $\dA$, $\dC$, $\dG$, $\dT$, and `blank'. The role of the blank symbols is critical, since without them, a pair of identical base predictions could indicate either a genuine repeated nucleotide or simply an unusually long dwell time of the same base in the pore. The use of blank symbols helps resolve this ambiguity. Another useful consequence of their inclusion is that the length of the predicted nucleotide sequence can be much shorter than a raw read, which is usually the case.

Thus, the probability matrix generated during CTC basecalling consists of a probability vector over the 5 states, one for each of the samples in a raw read. As done in case of the Lokatt basecaller, beam search exploits this probability matrix to explore the most likely state sequences that could have caused the observed raw signal.


\subsection{Primer Localization via \ps} \label{methods::primer-search}


Under the framework of the Lokatt basecaller, the task of detecting and localizing a specific target $k$-mer sequence $\boldsymbol{k}^* = (k_1, \ldots, k_l)$ in a raw signal $\boldsymbol{x}$ requires the efficient computation of the likelihood that, starting from a specific sample in $\boldsymbol{x}$, say $x_i$, the nanopore encountered the $k$-mers in $\boldsymbol{k}^*$. This likelihood can be formally expressed as $\sum_{j=i+l}^{N}p(\boldsymbol{k}^*, \boldsymbol{x}_i^{j})$
, where the summation is over possible durations over which the target sequence might be expressed in the raw signal. Thus, the objective of \pslokatt~is to directly estimate the location of $\boldsymbol{k}^*$ in $\boldsymbol{x}$, rather than basecalling $\boldsymbol{x}$ first, by adapting the GMBS algorithm to obtain an approximation of this likelihood. To accomplish this efficiently, \pslokatt~parallelizes the computation of $\sum_{j=i+l}^{N}p(\boldsymbol{k}^*, \boldsymbol{x}_i^j)$ over multiple candidate starting samples $x_i$. Fig.~\ref{fig:primer_search} demonstrates how this is achieved, namely by initiating multiple beams, each starting from a unique sample, say $x_i$, and propagating them to explore the most probable sequences of $k$-mer dwelling and transition events that best explain the subsequent current samples $(x_{i+1}, x_{i+2}, \ldots)$. Since there may be multiple equally-likely distinct sequences of dwelling and transition events for the observations $(x_i,x_{i+1},\ldots)$, their respective path likelihoods need to be aggregated to reliably approximate $\sum_{j=i+l}^{N}p(\boldsymbol{k}^*, \boldsymbol{x}_i^j)$. This is addressed by merging two extended beams whenever they are found to indicate the same sequence of $k$-mers and correspond to the same number of raw samples, thereby performing a similar on-the-go marginalization that was employed by the original GMBS algorithm in~\cite{xuMarginalizedBeamSearch2024}. We limit the computational complexity of this algorithm by constraining the maximum number of parent beams to a user-defined parameter $W$---whenever the number of parent beams exceeds $W$, only the $W$-highest scoring ones are preserved for the next round of beam extensions. 

Since a $k$-mer typically dwells in the nanopore for roughly $10$ samples,\footnote{Assuming the sampling rate of $4$ kHz and the translocation speed of $400$--$450$ bases per second of the R9 and R10 chemistries.} 
it is reasonable to expect that in addition to the sample where the target sequence actually began, say $x_{i^*}$, its immediately neighboring samples $x_{i^*-1}, x_{i^*+1}, x_{i^*+2}, \ldots$ etc. will also be associated with high primer occurrence likelihoods. Moreover, constraining the number of parent beams to $W$ creates competition among beams representing different starting samples---it would be suboptimal to jointly initiate beams corresponding to consecutive starting samples, since if they represent starting samples close to $x_{i^*}$, they will compete strongly against each other, likely leading to a reduced exploration of good paths for all plausible starting samples and resulting in a poor approximation of the desired likelihood. Therefore, we initiate beams from starting samples that are sufficiently far from each other to ensure that a promising beam is likely to compete against less plausible beams. In our simulations on raw reads from nanopore sequencers based on R9.4.1 chemistry, we found a minimum separation of $50$ samples to be sufficient. 

Note that \pslokatt~can also be straightforwardly adapted for use in the framework of CTC basecallers like Bonito by simply altering the probability computation to function over the state space of CTC basecallers. We implemented both Lokatt- and CTC-based versions of \ps,~as we discuss in~Sec.~\ref{sse:results_primerseeker}. 

It is worth mentioning that although it makes intuitive sense that the task of detecting a target sequence in $\boldsymbol{x}$ should be easier than basecalling it, the procedure described above usually incurs a complexity that is at least $10$ times higher than that of basecalling due to two major reasons. Firstly, as hinted earlier, multiple consecutive samples of the raw signal qualify as valid starting positions of the target primer sequence, owing to the random dwell time of the first $k$-mer of $\boldsymbol{k}^*$. Computing $\sum_{j=i+l}^{N}p(\boldsymbol{k}^*, \boldsymbol{x}_i^{j})$ for each of these potential starting positions is very redundant, since their respective beams involve very similar computations, as one might infer from Fig.~\ref{fig:primer_search}. Secondly, \ps~performs computations similar to a restricted form of basecalling on each window of the raw signal, say from sample $i$ to sample $i+D-1$, where $i$ ranges from $1$ to $N$ (the length of the raw signal), and the window length $D$ represents the maximum number of samples that the target sequence spans in $\boldsymbol{x}$. As a consequence, for each raw sample \ps~performs  significantly more computations than basecalling, because a specific sample may be associated with any of the nucleotides in the target sequence.

To address this additional computational burden, \ps~incorporates a filtering mechanism to eliminate redundant computations due to the examination of neighboring starting positions. Given an integer $s$, which we refer to as the `subsampling factor', this technique selects the most promising candidate starting position from each window of $s$ samples in $\boldsymbol{x}$, thus reducing the time complexity of \ps~by approximately a factor of $s$. In this manner, adjusting $s$ allows \ps~to be as efficient as basecalling at a negligible loss in performance (Supplementary~\ref{supp:complexity-primer-search}).


A complete description of this optimized version of \ps~can be found in Supplementary~\ref{supp:optimized-psctc}.

\subsection{Decoding via \syndec} \label{methods::decoding}

Our workflow, depicted in Fig.~\ref{fig:pipeline}, begins by using \ps~to identify the position of the starting primer. We then crop the two-dimensional probability matrix $\boldsymbol{P}$ 
generated by the neural network (see Sec.~\ref{sec:basecalling-beam-search}), ensuring that the first column of the new matrix aligns with the estimated primer position, and that the decoder of the convolutional code only examines the part of the raw signal that contains the codeword and its surrounding primers. 
Subsequently, this cropped matrix is fed to \syndec, which deploys a beam search algorithm that permits transitions to multiple bases as dictated by the quaternary syndrome trellis (this is in contrast to \ps, where only the transitions that correspond to the specified primer are allowed). Fig.~\ref{fig:beam-search-decoder} illustrates this workflow for \syndeclokatt; \syndecctc~operates along the same principles.

Note that our complete decoding pipeline, as depicted in Fig.~\ref{fig:pipeline}, can be seen as a modification of standard basecalling workflows. However, unlike traditional basecallers, \syndec~operates in a smaller search space, focusing exclusively on sequences that correspond to valid codewords.

In addition to using syndrome trellises to integrate the basecaller and the decoder, our decoding workflow differs from \chandakdec~in that it incurs time and space complexities independent of encoder memory and linear in payload length. Furthermore, our decoder, \syndec, starts and terminates with the flanking primers instead of just processing the part of the raw signal that contains the payload.  
This helps reduce errors caused by noise at the primer-payload boundaries.

\backmatter

\bmhead{Supplementary information}


A supplementary file is provided containing details of \ps, our simulation methodology, the encoding schemes used, additional experimental results, and complexity analysis.

\bmhead{Data \& Code Availability} \label{sec:data-code}


The raw FAST5 data, simulation scripts, and results are collated in 10.5281/zenodo.19158795. The GitHub repositories containing the source code for \ps~and \syndec, as well as a modified version of \chandakdec~supporting evaluation over additional datasets, are also referenced therein.

The raw FAST5 data used to generate the results in Fig.~\ref{fig:primer-search-sims} and Fig.~\ref{fig:fer-vs-discarded-reads} was obtained from the dataset by Volkel \textit{et al.} \cite{volkel_nanopore_2025}, which is available at 10.5281/zenodo.11985454. This dataset contains raw reads that were obtained by sequencing custom-designed oligos of length between $1250$--$5000$ bases, using ONT nanopores of flow cell chemistry R9.4.1. We also present results on two additional datasets in Supplementary~\ref{supp:decoding-sims}.

\bmhead{Acknowledgements}

The authors gratefully acknowledge financial support from the European Union (DiDaX 101115134, DNAMIC 101115389, NEO 101115317). R.S. received financial support from the Theory@EMBL Transversal Theme visitor programme. A.G. was partially supported by  the Swedish Research Council
(VR) under grant 2023-05065.












\bibliography{resources/literature}

\clearpage
 \begin{titlepage}
        \centering
        
        {\LARGE\bfseries Supplementary Materials for\par}
        \vspace{1.5em}
        
        {\Large SynDe: Syndrome--guided Decoding of Raw Nanopore Reads \par} 
        \vspace{1.5em}
        
        {\large Anisha Banerjee$^{1}$, Roman Sokolovskii$^{2,3}$, Thomas Heinis$^{2}$, Antonia Wachter-Zeh$^{1}$, Eirik Rosnes$^{4}$, Alexandre Graell i Amat$^{5}$ \par} 
        \vspace{0.5em}
        {\small 
          $^{1}$Institute for Communications Engineering; Technical University of Munich (TUM), Munich, Germany. \par
          $^{2}$Department of Computing, Imperial College London, UK. \par
          $^{3}$European Molecular Biology Laboratory, European Bioinformatics Institute (EMBL-EBI), Wellcome Genome Campus, Hinxton, UK. \par
          $^{4}$Simula UiB, Bergen, Norway. \par
          $^{5}$Department of Electrical Engineering, Chalmers University of Technology, Gothenburg, Sweden. \par         
        }
        \vspace{1em}
        {\normalsize Correspondence to: anisha.banerjee@tum.de \par}
        \vspace{2em}
        
        \raggedright
        {\bfseries This PDF file includes:}\par
        Supplementary Text

        Supplementary Figures S1 to S5.

        Supplementary Tables S1 to S4.
        
        \vspace{1em}
        
        {\bfseries Other Supplementary Materials for this manuscript include the following:}\par

        Source code and simulation scripts are available at: \url{https://zenodo.org/records/19158795}; \url{https://github.com/anisha-ban/decoding-raw-nanopore-signals-dna-storage}
        
        \vfill
    \end{titlepage}

\appendix

\renewcommand{\thesection}{S\arabic{section}}

\counterwithout{figure}{section}
\setcounter{figure}{0}
\renewcommand{\thefigure}{S\arabic{figure}}

\counterwithout{table}{section}
\setcounter{table}{0}
\renewcommand{\thetable}{S\arabic{table}}
\makeatletter
\renewcommand{\p@table}{}
\renewcommand{\fnum@table}{\tablename~\thetable} 
\makeatother


\begin{center}
    {\Large\bfseries Supplementary Text}
\end{center}

\section{Description of \ps} \label{supp:optimized-psctc}

In this section, we provide an algorithmic description of \ps, the pseudocode of which can be found in Algorithm~\ref{alg:ps-opt}. As Fig.~\ref{fig:pipeline} indicates, \ps~requires as input the probability matrix $\boldsymbol{P}$ produced by the neural network and the target sequence $\boldsymbol{k}$ to be sought in the raw signal (or more accurately, in the matrix $\boldsymbol{P}$). 

\begin{algorithm}[h]
\caption{\ps}\label{alg:ps-opt}


\KwInput{Probability matrix $\boldsymbol{P} \in \mathbb{R}^{Q \times N}$, target sequence $\boldsymbol{k}$}
\KwParam {Subsampling factor $s$, number of beams $W$, extension depth $d$, maximum depth $d_{\textnormal{max}}$, shift $\delta$, and concentration threshold $\tau$}

\KwOutput{Estimated starting position of $\boldsymbol{k}$ in raw signal $t^*$}

\DontPrintSemicolon

$t^* \gets -1$, $p^* \gets 0$ 

$N \gets \text{number of columns in }\boldsymbol{P}$ \tcp*{Usually equals the length of the raw signal}

$A \gets \min(W \cdot \delta \cdot s, \lceil N/2 \rceil)$ 

$t_{\textnormal{start}} \gets 1$, $t_{\textnormal{end}} \gets t_{\textnormal{start}} + A-1$ 

\tcc{Iterate over blocks of $A$ columns of matrix $\boldsymbol{P}$, continue until halfway through $\boldsymbol{P}$}

\While{$t_{\textnormal{start}} \leq \lceil N/2 \rceil$}{    
    $\mathcal{B} \gets \emptyset$ \tcp*{set of parent beams}
    
    \For{$i = t_{\textnormal{start}}$ \textbf{to} $t_{\textnormal{end}} - 1$}{
        $\boldsymbol{b} \gets \langle \texttt{prob} = 0, \texttt{start\_pos}=i, \texttt{end\_pos} = i, \texttt{sequence\_length} = 1 \rangle$ 

        $\boldsymbol{b}.\texttt{prob} \gets$ probability that the $i$-th sample of the raw signal\footnote{rather, the $i$-th column of the probability matrix $\boldsymbol{P}$} corresponds to the first unit of $\boldsymbol{k}$ 
        
        $\mathcal{B} \gets \{\boldsymbol{b}\} \cup \mathcal{B}$ 
    }
    
    \If{$s > 1$}{
        $\mathcal{B} \gets \textsc{SubsamplingBeamSearch}(\mathcal{B}, \boldsymbol{P}[:,t_{\textnormal{start}}:t_{\textnormal{end}}], \boldsymbol{k}, s, d)$
    }

   $\mathcal{B} \gets \textsc{Prune}(\mathcal{B}, \tau)$ \tcp*{Keep top beams comprising $\tau$ fraction of total probability}

    $t', p'\gets\textsc{PrimerBeamSearch}(\mathcal{B}, \boldsymbol{P}[:,t_{\textnormal{start}}:t_{\textnormal{end}}], \boldsymbol{k}, d_{\textnormal{max}}, \delta)$ 

    \If{$p' > p^*$}{
        $p^* \gets p'$, $t^* \gets t' + t_{\textnormal{start}} - 1$ 
    }

    $t_{\textnormal{start}} \gets t_{\textnormal{end}}$, $t_{\textnormal{end}} \gets \min(t_{\textnormal{start}} + A, \lceil N/2 \rceil)$ 
}
\KwRet{$t^*$}
\end{algorithm}

\ps~attempts to find the sample of the raw signal that maximizes the probability that the upcoming raw samples (columns of $\boldsymbol{P}$) correspond to the target sequence $\boldsymbol{k}$. It does so by considering batches of $A$ consecutive candidate starting positions, and performing a beam search operation on each of them.

For our purposes, a beam is essentially a data structure that encapsulates the following variables: 
\begin{itemize}
    \item \texttt{start\_pos} and \texttt{end\_pos} denote the starting and ending column indices spanned by the beam.
    \item \texttt{sequence\_length} refers to the length of the prefix of $\boldsymbol{k}$ that the beam has covered thus far.
    \item \texttt{prob} indicates the total probability of the beam, i.e., the probability that the samples of the raw signal (or columns of $\boldsymbol{P}$) indexed from \texttt{start\_pos} to \texttt{end\_pos}, correspond to the first \texttt{sequence\_length} units of $\boldsymbol{k}$.
\end{itemize}

For the list of candidate starting positions $\{t, t+1, \ldots, t+A-1\}$, \ps~begins by creating $A$ beams, each of which represents a unique starting position from the list through the member variable \texttt{start\_pos}. Initially, \texttt{end\_pos} is set equal to \texttt{start\_pos} and \texttt{sequence\_length} is set to $1$, while \texttt{prob} is assigned $\boldsymbol{P}[\boldsymbol{k}[1], \texttt{start\_pos}]$, i.e., the probability that the \texttt{start\_pos}-th raw sample (or column of $\boldsymbol{P}$) is associated with the first entry of $\boldsymbol{k}$. Subsequently, this initial set of $A$ beams, say $\mathcal{B}$, may be optionally subjected to a pruning step, based on the user-defined parameter $s$. This component of the algorithm, elaborated in Algorithm~\ref{alg:subsampling-beam-search}, effectively seeks to filter out the most promising candidate in each neighborhood of $s$ starting positions, in an effort to address the issue of repeated computations discussed in Section~\ref{methods::primer-search}. It involves splitting the set $\mathcal{B}$ into $s$ different buckets, say $\mathcal{B}_0, \ldots, \mathcal{B}_{s-1}$, such that the $j$-th bucket $\mathcal{B}_j$ contains only the beams in $\mathcal{B}$ whose starting positions satisfy $\texttt{start\_pos} \bmod s \equiv j$. Following this, for a user-specified integer $d > 1$ and $j \in \{0,\ldots,s-1\}$, the beams in $\mathcal{B}_j$ undergoes $s-j+d$ iterations of propagation, merging, sorting, and pruning, identically to Fig.~\ref{fig:primer_search}. In each such iteration, the variable \texttt{end\_pos} of the involved beams is incremented by one. This means that after the completion of this staggered beam propagation process, the beams in bucket $\mathcal{B}_j$ will satisfy $\texttt{end\_pos} = \texttt{start\_pos} + s  - j + d$, or equivalently, $\texttt{end\_pos} \bmod s \equiv d$, which is notably independent of $j$. In other words, we propagate beams from different starting positions to different depths such that their end positions become aligned, which allows subsequent merging, as described below.

The beams in all buckets $\mathcal{B}_0, \ldots,\mathcal{B}_{s-1}$ are then combined into a new set $\mathcal{B}'$ and subjected to a different kind of merging operation---namely, that if any two distinct beams $\boldsymbol{b}_1, \boldsymbol{b}_2 \in \mathcal{B}'$ match in their values of \texttt{end\_pos} and \texttt{sequence\_length}, respectively, then they are unified into a single beam $\boldsymbol{b}'$, with \texttt{prob} set to $\boldsymbol{b}_1.\texttt{prob}+\boldsymbol{b}_2.\texttt{prob}$ and \texttt{start\_pos} is assigned the corresponding variable of the beam with higher probability. 

\begin{algorithm}
\caption{\textsc{SubsamplingBeamSearch}}\label{alg:subsampling-beam-search}

\DontPrintSemicolon

\KwIn{Set of beams $\mathcal{B}$, probability matrix $\boldsymbol{P} \in \mathbb{R}^{Q \times N}$, target sequence $\boldsymbol{k}$, and subsampling factor $s$}
\KwParam{Extension depth $d$}
\KwOut{Pruned set of beams $\mathcal{B}'$}

$W \gets (\text{number of beams in } \mathcal{B}) / s$ \tcp*{Target number of beams}

$N \gets \text{number of columns in } \boldsymbol{P}$ \tcp*{Length of raw signal}

\For{$i = 0$ \KwTo $s - 1$}{
    $\mathcal{B}_i \gets \{\boldsymbol{b} \in \mathcal{B} : \boldsymbol{b}.\texttt{start\_pos} \bmod s \equiv i\}$ \tcp*{$i$-th bucket}
    $n_{\textnormal{steps}} \gets s - i + d$ \tcp*{Number of propagation steps for beams}
    
    \For{$t = 1$ \KwTo $n_{\textnormal{steps}}$}{
        $\mathcal{V} \gets \emptyset$ \tcp*{Initialize set of child beams}
        
        \ForEach{$\boldsymbol{b} \in \mathcal{B}_i$ such that $\boldsymbol{b}.\texttt{end\_pos} + 1 < N$}{
            
            $p' \gets \boldsymbol{b}.\texttt{prob} \cdot \boldsymbol{P}[\boldsymbol{k}[\boldsymbol{b}.\texttt{sequence\_length}], \boldsymbol{b}.\texttt{end\_pos}+1]$ \;
            
            $\boldsymbol{b}_{\textnormal{dwell}} \gets \langle \texttt{prob}=p', \texttt{start\_pos}=\boldsymbol{b}.\texttt{start\_pos}, \texttt{end\_pos}=\boldsymbol{b}.\texttt{end\_pos}+1,\texttt{sequence\_length}=\boldsymbol{b}.\texttt{sequence\_length} \rangle$ \;

            $\hat{p} \gets \boldsymbol{b}.\texttt{prob} \cdot \boldsymbol{P}[\boldsymbol{k}[\boldsymbol{b}.\texttt{sequence\_length}+1], \boldsymbol{b}.\texttt{end\_pos}+1]$ \;

            $\boldsymbol{b}_{\textnormal{ext}}\gets \langle \texttt{prob}=\hat{p}, \texttt{start\_pos}=\boldsymbol{b}.\texttt{start\_pos}, \texttt{end\_pos}=\boldsymbol{b}.\texttt{end\_pos}+1,\texttt{sequence\_length}=\boldsymbol{b}.\texttt{sequence\_length}+1 \rangle$ \;
                
            $\mathcal{V} \gets \mathcal{V} \cup \{\boldsymbol{b}_{\textnormal{dwell}}, \boldsymbol{b}_{\textnormal{ext}}\}$
            
        }
        $\mathcal{B}_i \gets \mathcal{V}$ \;
        
        \If{$\mathcal{B}_i = \emptyset$}{
            \textbf{break}
        }
        
        \tcc{Merge beams with same starting position and sequence length}
        \ForEach{pair of distinct beams $\boldsymbol{b}_1, \boldsymbol{b}_2 \in \mathcal{B}_i$ such that $\boldsymbol{b}_1.\texttt{start\_pos} = \boldsymbol{b}_2.\texttt{start\_pos}$ \textbf{and} $\boldsymbol{b}_1.\texttt{sequence\_length} = \boldsymbol{b}_2.\texttt{sequence\_length}$ }{
            $\boldsymbol{b}_{1}.\texttt{prob} \gets \boldsymbol{b}_{1}.\texttt{prob} + \boldsymbol{b}_{2}.\texttt{prob}$ \;
        
            $\mathcal{B}_{i} \gets \mathcal{B}_{i} \setminus \{\boldsymbol{b}_2\}$ \;
        }
        
        Sort $\mathcal{B}_i$ in descending order of probability and only keep its top $W$ beams \;
        
    }
}

$\mathcal{B}' \gets \bigcup_{i=0}^{s-1} \mathcal{B}_i$ \;

\tcc{Merge beams in $\mathcal{B}'$ that end at same sample and have same sequence length}
\ForEach{pair of distinct beams $(\boldsymbol{b}_1, \boldsymbol{b}_2) \in \mathcal{B}'$ such that $\boldsymbol{b}_1.\texttt{end\_pos} = \boldsymbol{b}_2.\texttt{end\_pos}$ \textbf{and} $\boldsymbol{b}_2.\texttt{sequence\_length} = \boldsymbol{b}_2.\texttt{sequence\_length}$}{
    {
        $\boldsymbol{b}_{1}.\texttt{prob} \gets \boldsymbol{b}_{1}.\texttt{prob} + \boldsymbol{b}_{2}.\texttt{prob}$ \;
            
        $\mathcal{B}' \gets \mathcal{B}' \setminus \{\boldsymbol{b}_2\}$ \;
    }
}

Sort $\mathcal{B}'$ in descending order of probability and only keep its top $W$ beams

\KwRet{$\mathcal{B}'$}
\end{algorithm}

This merging step is followed by the usual sorting operation which orders the beams in $\mathcal{B}'$ is descending order of their respective \texttt{prob} values, and the top $A/s$ number of beams of $\mathcal{B}'$ are preserved, while the rest are discarded. Optionally, $\mathcal{B}'$ may be subjected to a further pruning round, the extent of which is controlled by the user-defined fraction $\tau$, referred to as the concentration threshold. Specifically, this step involves only preserving the top $Z<|\mathcal{B}'|$ beams of $\mathcal{B}'$, such that the sum of probabilities of these surviving $Z$ beams is at most $\tau$ times the sum of probabilities of the beams in the initial version $\mathcal{B}'$. 

The final stage of \ps~constitutes subjecting $\mathcal{B}'$ to the familiar beam propagation process explained in Fig.~\ref{fig:primer_search} and laid out more formally in Algorithm~\ref{alg:primer-beam-search}.

In our simulations on Lau \textit{et al.}'s dataset, we observed that as the subsampling factor $s$ increases from $1$, the runtime complexity of \psctc~decreases linearly at first. However, the rate of this decrease soon starts falling, before eventually reaching a plateau at $s=8$. This speedup is usually accompanied by only a slight degradation in performance of \ps ---a less than $1\%$ drop in the fraction of reads for which the estimates from \psctc~and Chandak \textit{et al.}'s approach differ by no more than $50$. By cleverly tuning the subsampling factor $s$, the number of beams $W$, and the concentration threshold $\tau$, we demonstrate on the dataset by Lau \textit{et al.} that \psctc~achieves a time complexity similar to that of basecalling (Supplementary~\ref{supp:complexity-primer-search}).

\begin{algorithm}
\caption{\textsc{PrimerBeamSearch}}\label{alg:primer-beam-search}
\DontPrintSemicolon
\KwInput{Set of beams $\mathcal{B}$, probability matrix $\boldsymbol{P} \in \mathbb{R}^{Q \times N}$, target sequence $\boldsymbol{k}$, and shift $\delta$}
\KwOutput{Best starting position $t^*$, best score $p^*$}
$t^* \gets -1$, $p^* \gets 0$\; 
$N \gets \text{number of columns in }\boldsymbol{P}$\; 
$\boldsymbol{\Phi} \gets [0.0, \ldots, 0.0]$ of length $N$ \;
$L \gets \text{length of } \boldsymbol{k}$ \;
\tcc{Distribute beams into buckets based on starting position}
\For{$j = 0$ \textbf{to} $\delta - 1$}{
    $\mathcal{B}_j \gets \{\boldsymbol{b} \in \mathcal{B} : \boldsymbol{b}.\texttt{start\_pos} \bmod \delta \equiv j\}$, $\phi_j \gets \sum_{\boldsymbol{b} \in \mathcal{B}_j} \boldsymbol{b}.\texttt{prob}$ \;
}
\tcc{Start beam propagation process}
\While{$\max_{j}\phi_j > 0$}{
    $j^* \gets \arg\max_{j} \phi_j$\; 
    $\mathcal{V} \gets \emptyset$ \;
    
    \ForEach{$\boldsymbol{b} \in \mathcal{B}_{j^*}$ such that $\boldsymbol{b}.\texttt{prob} \geq p^*$, $\boldsymbol{b}.\texttt{sequence\_length} \leq L$, and $\boldsymbol{b}.\texttt{end\_pos} < N$}{
        $t_{\textnormal{start}} \gets \boldsymbol{b}.\texttt{start\_pos}$, $t_{\textnormal{end}} \gets \boldsymbol{b}.\texttt{end\_pos}$\;
        $\ell \gets \boldsymbol{b}.\texttt{sequence\_length}$ \;
                
        $p' \gets \boldsymbol{b}.\texttt{prob} \cdot \boldsymbol{P}[{\boldsymbol{k}[\ell], t_{\textnormal{end}}}+1]$ \tcp*{Dwelling probability}
        $\boldsymbol{b}_{\textnormal{dwell}} \gets \langle \texttt{prob}=p', \texttt{start\_pos}=t_{\textnormal{start}}, \texttt{end\_pos}=t_{\textnormal{end}}+1,\texttt{sequence\_length}=\ell+1 \rangle$ \;
        
        $\hat{p} \gets \boldsymbol{b}.\texttt{prob} \cdot \boldsymbol{P}[{\boldsymbol{k}[\ell+1], t_{\textnormal{end}}+1}]$ \tcp*{Extension probability}
        $\boldsymbol{b}_{\textnormal{ext}} \gets \langle \texttt{prob}=\hat{p}, \texttt{start\_pos}=t_{\textnormal{start}}, \texttt{end\_pos}=t_{\textnormal{end}}+1,\texttt{sequence\_length}=\ell+1 \rangle$ \;
        
        $\mathcal{V} \gets \mathcal{V} \cup \{\boldsymbol{b}_{\textnormal{dwell}}, \boldsymbol{b}_{\textnormal{ext}}\}$ \;
    }
    $\mathcal{B}_{j^*} \gets \mathcal{V}$ \;
    
    \lIf{$\mathcal{B}_{j^*} = \emptyset$}{$\phi_{j^*} \gets 0$, \textbf{continue}}
    
    \tcc{Merge beams with same starting position and sequence length}
    \ForEach{pair of distinct beams $\boldsymbol{b}_1, \boldsymbol{b}_2 \in \mathcal{B}_{j^*}$ such that $\boldsymbol{b}_{1}.\texttt{start\_pos} = \boldsymbol{b}_{2}.\texttt{start\_pos}$ \textbf{and} $\boldsymbol{b}_{1}.\texttt{sequence\_length} = \boldsymbol{b}_{2}.\texttt{sequence\_length}$}{
        
        $\boldsymbol{b}_{1}.\texttt{prob} \gets \boldsymbol{b}_{1}.\texttt{prob} + \boldsymbol{b}_{2}.\texttt{prob}$\; 
        $\mathcal{B}_{j*} \gets \mathcal{B}_{j^*} \setminus \{\boldsymbol{b}_2\}$ \;
    }
    Sort $\mathcal{B}_{j*}$ in descending order of probability and only keep its top $W$ beams \;
    $\phi_{j^*} \gets \sum_{\boldsymbol{b} \in \mathcal{B}_{j^*}} \boldsymbol{b}.\texttt{prob}$ \;
    
    \ForEach{beam $\boldsymbol{b} \in \mathcal{B}_{j^*}$ such that $\boldsymbol{b}.\texttt{sequence\_length} = L$}{
        $\boldsymbol{\Phi}[\boldsymbol{b}.\texttt{start\_pos}] \gets \boldsymbol{\Phi}[\boldsymbol{b}.\texttt{start\_pos}] + \boldsymbol{b}.\texttt{prob}$ \;
        \If{$p^*  < \boldsymbol{\Phi}[\boldsymbol{b}.\texttt{start\_pos}] $}{$t^* \gets \boldsymbol{b}.\texttt{start\_pos}$, $p^* \gets \boldsymbol{\Phi}[\boldsymbol{b}.\texttt{start\_pos}]$ \;}
    }
}
\KwRet{$t^*$,$p^*$}
\end{algorithm}

\section{Simulations} \label{supp:decoding-sims}
 \label{supp:encoding}


Recall from Section~\ref{sec:conv-codes} that the parity-check matrix of a $(c,b,\nu)$ convolutional code of length $N$ consists of repeated instances of a smaller submatrix of $c$ columns. This means that the complete syndrome trellis (of $N+1$ levels) can be derived from a smaller syndrome trellis (of $c+1$ levels) corresponding to the aforementioned smaller parity-check matrix. Thus, for the purpose of our decoding simulations, we only need to store a representation of this smaller syndrome trellis in the cache memory. 
As mentioned earlier, this syndrome trellis usually has $2^{c-b+\nu}$ syndrome states at each level. 

For fast access, we store the minimal syndrome trellis in cache memory by saving the numerical representation of each syndrome state as a $32$-bit variable, in addition to using $4$ extra bits for each of the $4$ possible DNA symbols, whether or not it can succeed a specific syndrome state. Thus, the total space necessary to store the minimal trellis amounts to at most $2^{c-b+\nu} \cdot c \cdot (32+4)$ bits. Assuming that the cache memory can hold up to $32$MB and that $c\leq 20$, this implies that convolutional codes with high rate (small value of $c-b$) and memory up to $\nu = 18$ can be practically used with \syndec. For a general linear code of $240$ bits (equivalently, $120$ quaternary symbols) and rate $r$, each level of the associated syndrome trellis has at most $2^{(1-r)\cdot 240}$ states. Using equivalent arguments as before, we conclude that \syndec~can accommodate such codes provided that $r \geq 0.934$.

The codes we use in our simulations are listed in Table~\ref{tab:codes1}, while  in Table~\ref{tab:codes2} we list the codes used to evaluate \chandakdec. While the syndrome trellis complexity of the codes in Table~\ref{tab:codes1} is dictated by $\nu$, \chandakdec~operates on a different kind of trellis where each state corresponds to the content of the $\nu$ shift registers in the encoder of a convolutional code. Thus, the number of states at each level is given by $2^{\nu}$, meaning that $\nu$ also dictates the trellis complexity in the case of \chandakdec.

\begin{table}[h]
    \centering
    \scalebox{0.85}{\begin{tabular}{c c c c c c c c}
    \toprule
    Identifier & \shortstack{\!Convolutional code\!\\$(c,b,\nu)$} &
    \shortstack{Marker\\period} &
    \shortstack{Message\\length (bits)}  &
    \shortstack{Payload\\length} & Rate & \shortstack{Mean beam complexity of \\ \!\syndecctc~on Lau \textit{et al.}'s dataset\!} \\
    \midrule
    CCM10-7 & $(13,12,10)$ & $7$ &  $170$ &   $126$ & $0.762$ & $1.84\times 10^3$ \\
    CCM10-6 & $(13,12,10)$ & $6$ & $166$ & $129$ & $0.744$ & $1.83 \times 10^3$ \\
    CCM10-5 & $(13,12,10)$ & $5$ & $140$ & $117$ & $0.718$ & $1.78\times 10^3$ \\
    CCM9-5 & $(9,8,9)$ & $5$ & $134$ &  $114$ & $0.702$ & $1.73 \times 10^3$ \\
    CCM10-4 & $(11,10,10)$ & $4$ & $112$ &  $106$ & $0.68$ & $1.78 \times 10^3$ \\
    CCM9-14 & $(4,3,9)$ & $14$ & 150 & $113$ & $0.664$ & $2.05\times 10^3$ \\
    \bottomrule
\end{tabular}}
    \caption{Summary of coding schemes used for \syndec. 
    The convolutional codes were designed in \cite{rosnesMaximumLengthConvolutional2004}. These combined schemes of convolutional codes and marker codes are assigned identifiers to indicate their respective marker periods and parameter value $\nu$. Specifically, a code CCMx-y corresponds to a code construction where the convolutional code has $\nu=\textnormal{x}$ and the marker code constitutes a single marker symbol inserted after every y symbols of the convolutional codeword. }
    \label{tab:codes1}
\end{table}
\begin{table}[h]
    \centering
    \scalebox{0.85}{\begin{tabular}{c c c c c c c c}
    \toprule
    Identifier &
    \shortstack{Convolutional code\\$(c,b,\nu)$} &
    \shortstack{Message\\length (bits)} &
    Payload length &
    Rate & \shortstack{Mean beam complexity of\\ \chandakdec~on Lau \textit{et al.}'s dataset} \\
    \midrule
    CC6-5 & $(6,5,6)$ & $172$ & $112$ & $0.768$ & $1.8 \times 10^5$ \\
    CC8-5 & $(6,5,8)$ & $172$ &  $113$ & $0.761$ & $7.3 \times 10^5$ \\
    CC11-5 & $(6,5,11)$ & $172$ & $115$ & $0.7478$ & $5.7 \times 10^6$ \\
    CC6-3 & $(4,3,6)$  & $157$ & $114$  & $0.688$ & $1.7 \times 10^5$\\
    CC8-3 & $(4,3,8)$ & $156$ & $115$ & $0.678$ & $6.8 \times 10^5$ \\
    CC11-3 & $(4,3,11)$ & $156$ & $117$ & $0.666$ & $5.2\times 10^6$\\
    \bottomrule
\end{tabular}}
    \caption{Summary of coding schemes used in \cite{lauMagneticDNARandom2023, chandakOvercomingHighNanopore2020}. 
    These codes were designed in \cite{livaCodeDesignShort2016}, and we assign them identifiers to indicate their respective parameter values $\nu$ and $b$, i.e., a code CCx-y has $\nu=\textnormal{x}$ and $b=\textnormal{y}$. The rate values stated here account for the $8$-bit CRC that is appended to the message vector prior to convolutional encoding in Chandak \textit{et al.}'s pipeline.   }
    \label{tab:codes2}
\end{table}

To test these encoding schemes on raw nanopore sequencing reads, we first map every raw read to its reference sequence via basecalling and alignment. Subsequently, we conduct simulations (of \ps~and \syndec) by assuming that a portion of the reference sequence corresponds to a random codeword, translated by a pseudo-random offset known to the decoder.  We consider payload sequences that are between $105$ and $130$ bases in length, with primer sequences of $25$ bases\footnote{The last $23$ bases of the primers are considered on Lau \textit{et al.}'s dataset.} on each side.

When using \ps, we maintained a minimum separation of $100$ samples (or $100/\sigma$ if the accompanying neural network has a stride of $\sigma$ samples.). In particular, \pslokatt~was run with $W=512$ and did not employ any runtime optimization tricks, i.e., it was used with $s=1$ and $\tau=1.0$. On the contrary, \psctc~used the parameter values $W=8$, $s=6$, and $\tau=0.98$. However, both implementations of \syndec~were run with $W=512$. \syndecctc~is built on the Bonito basecaller (v0.1.2), and the neural network model used is \texttt{dna{\textunderscore}r9.4.1}.

When decoding with \syndec, the generated likelihood scores are used to judge the reliability of the decoding result. In particular, instead of using CRC checks as in \chandakdec, our decoding scheme utilizes a user-defined threshold to decide whether a specific decoding result should be accepted, i.e., if the likelihood score of the output codeword exceeds the threshold, it is accepted. To produce  FER performance curves similar to Fig.~\ref{fig:fer-vs-discarded-reads}, we vary this user-defined threshold, and for each specific threshold, we measure the percentage of raw reads for which \syndec's output scores exceeded the threshold, as well as the FER of these remaining reads.

\subsection{Performance of \ps}

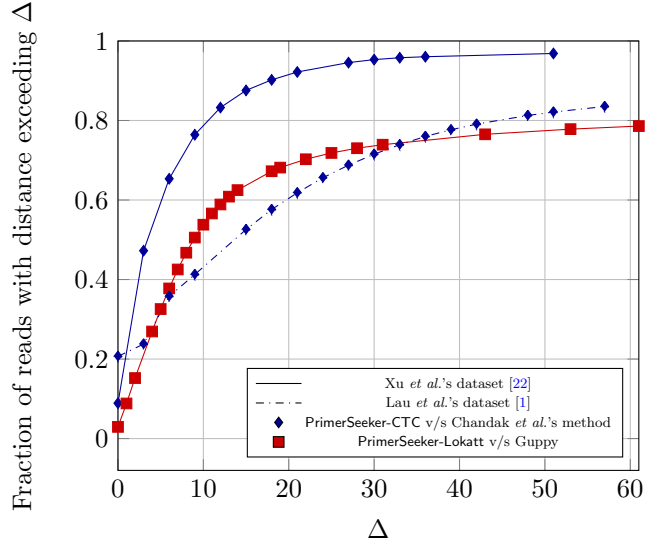
\begin{figure}
    \centering
  \begin{tikzpicture}
      \begin{axis}[legend style={nodes={scale=0.6, transform shape}},
				xmin=0, xmax=61,
				ymin=-0.08, ymax=1.0,
                xtick={0,10,20,30,40,50,60},
				xlabel=\textsc{$\Delta$}, ylabel={Fraction of reads with distance exceeding $\Delta$},
				legend pos=south east, legend style={font=\scriptsize},
				grid=both, 
                legend entries={Xu \textit{et al.}'s dataset \cite{xuLokattHybridDNA2023},
                Lau \textit{et al.}'s dataset \cite{lauMagneticDNARandom2023},
                \psctc~v/s Chandak \textit{et al.}'s method,
                \pslokatt~v/s Guppy},]
            \addlegendimage{no markers,solid}
            \addlegendimage{no markers,dash dot}
            \addlegendimage{only marks, mark=diamond*, mark options={fill=blue!60!black}}
            \addlegendimage{only marks, mark=square*, mark options={fill=red!80!black}}
            \addplot[solid, blue!60!black, mark=diamond*, mark options={fill=blue!60!black}, restrict x to domain=0:61] table {resources/primer-search-plots/primerseeker_xu_ctc_vs_chandak.txt};
            
            \addplot[solid, red!80!black, mark=square*, mark options={fill=red!80!black}, restrict x to domain=0:61] table {resources/primer-search-plots/primerseeker_xu_lokatt_vs_guppy.txt};
            
            \addplot[dash dot, blue!60!black, mark=diamond*, mark options={fill=blue!60!black}, restrict x to domain=0:61] table {resources/primer-search-plots/primerseeker_lau_ctc_vs_chandak.txt};

        \end{axis}
  \end{tikzpicture}
    \caption{Performance of the primer search algorithm on the datasets from Xu \textit{et al.} and Lau \textit{et al.} (see Table~\ref{tab:datasets}). Similar to Fig.~\ref{fig:primer-search-sims}, we observe that on both datasets, the estimates obtained from \psctc~and Chandak \textit{et al.}'s method are very close for most of the raw reads. The performance on the latter dataset is worse, especially so when we compare \pslokatt~with the Guppy-based technique. This is most likely owed to the fact that Lau \textit{et al.}'s dataset involves much shorter reads, on which basecalling performance is usually poorer. 
    }
    \label{supp:fig:primer-search-sims}
\end{figure}

Similar to our results in Fig.~\ref{fig:primer-search-sims}, we present the performance of \psctc~and \pslokatt~on the datasets from Lau \textit{et al.} and Xu \textit{et al.} (Table~\ref{tab:datasets}) in Fig.~\ref{supp:fig:primer-search-sims}.

In particular, we consider $24300$ raw nanopore signals from the dataset from Lau \textit{et al.} and observe that for 79.54\% of these reads, \psctc~offers estimates that are within $30$ samples of the estimates generated by the method employed in Lau \textit{et al.}. Again, on $95.31\%$ of $9600$ raw reads sampled from the dataset by Xu \textit{et al.}, the estimates offered by \psctc~and~Lau \textit{et al.}'s approach are within $30$ samples of each other. 

While \ps~is tasked with detecting target sequences of $25$ nucleotides on the datasets from Xu \textit{et al.} and Volkel \textit{et al.}, we choose to ignore the first $2$ nucleotides of the true forward primer sequence on Lau \textit{et al.}'s dataset, i.e., the sought target sequence is the length-$23$ suffix of the actual flanking sequence. As mentioned in Section~\ref{sec:discussion}, doing so not only bypasses potential boundary artifacts in the raw signal but also provides robustness against neural network errors in this region.



\begin{table}[!t] 
    \centering
    \scalebox{1.0}{\begin{tabularx}{\textwidth}{c p{5cm} p{5cm}}
    \toprule
    Source & Repository & Description \\
    \midrule
    Volkel \textit{et al.} \cite{volkel_nanopore_2025} & 10.5281/zenodo.11985454 & Custom-designed oligos of length between $1250$--$5000$ bases, each with a homopolymer length at most $3$ and with GC content of $50\%$ over $12$ base-pair windows. \\
    Xu \textit{et al.} \cite{xuLokattHybridDNA2023} & 10.5281/zenodo.7995806  &  Ecoli non-methylated genomic DNA (D5016, Zymo Research). \\
    Lau \textit{et al.} \cite{lauMagneticDNARandom2023} & \url{https://github.com/shubhamchandak94/nanopore_dna_storage_data/tree/bonito} & $15000$ oligos, each carrying a payload of $108$--$119$ bases, flanked by primers of $25$ bases on both sides. \\
    \bottomrule
\end{tabularx}}
    \caption{Nanopore sequencing datasets used for the simulations in this work. All used ONT MinION devices with flow cell chemistry R9.4.1.}
    \label{tab:datasets}
\end{table}

\subsection{Improvement with Marker Codes} \label{supp::sec:marker-codes}

\begin{figure} 
    \centering
    \scalebox{1.2}{\begin{tikzpicture}
      \begin{axis}[legend style={nodes={scale=0.55, transform shape}},
				ymode=log,
				xmin=-7, xmax=105,
				ymin=1.001e-3, ymax=1.0,
				xlabel={\small Percentage of reads discarded (\%)}, ylabel={\small Frame Error Rate},
				legend pos=south west, legend style={font=\scriptsize},
				grid=both]
            \addplot[dash dot, red!90!black, mark=*, mark options={fill=red!90!black}] table {resources/fer-discarded-reads-plots/volkel_dataset/FER_CTC_cc9_6_4_r0.643.txt};
            \addlegendentry{CC4-6, rate=$0.643$};
            \addplot[dash dot, blue!90!black, mark=*, mark options={fill=blue!90!black}] table {resources/fer-discarded-reads-plots/volkel_dataset/FER_CTC_cc10_7_5_r0.675.txt};
            \addlegendentry{CC5-7, rate=$0.675$};
            \addplot[solid, green!40!black, mark=*, mark options={fill=green!40!black}] table {resources/fer-discarded-reads-plots/volkel_dataset/FER_CTC_cc13_12_10_m5_r0.718.txt};
            \addlegendentry{CCM10-5, rate=$0.718$};
            \addplot[solid, blue!90!black, mark=*, mark options={fill=blue!90!black}] table {resources/fer-discarded-reads-plots/volkel_dataset/FER_CTC_cc11_10_10_m4_r0.68.txt};
            \addlegendentry{CCM10-4, rate=$0.68$};
        \end{axis}
  \end{tikzpicture}}
    \caption{Examination of how the decoding performance of \syndecctc~changes with the incorporation of marker symbols into convolutional codes on the dataset from Volkel \textit{et al.}.  }\label{supp:fig:fer-vs-discarded-reads-markers-comparison-volkel}
\end{figure}
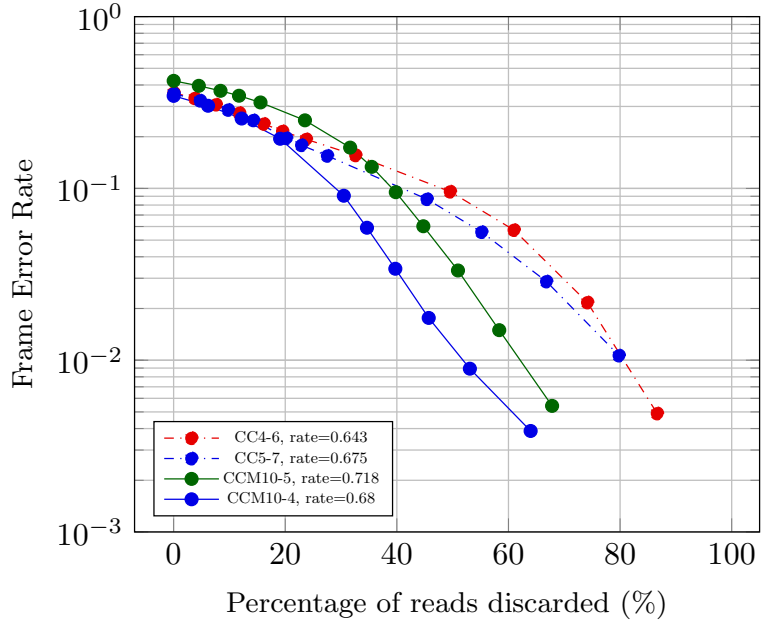

Fig.~\ref{supp:fig:fer-vs-discarded-reads-markers-comparison-volkel} illustrates the performance of \syndecctc~across four coding schemes. Two of these schemes just employ convolutional codes, while the other two incorporate constant marker symbols into the convolutional codes. Although the latter schemes have a higher rate than the convolutional codes, they provide a better trade-off between the fraction of discarded reads and the FER. This may seem counterintuitive at first, but it is important to consider that, unlike the binary symmetric channel, the varying dwell times of different $k$-mers in nanopore sequencers make it challenging to accurately determine the codeword position to which a specific sample in the raw signal corresponds, as discussed in Section~\ref{sec:conv-codes}. The addition of marker symbols helps mitigate this uncertainty to some extent, leading to an improved FER performance.

\subsection{\syndec: Motivation and Performance} \label{supp::sec::synde-perf}

The primary idea behind \syndec~is similar to the motivation behind sequential decoding of convolutional codes. That is, if the channel is perfectly noiseless, we need not use trellis-based decoders (Viterbi decoders) since a beam search decoder with a single beam ($W=1$) will offer an identical FER performance. On the other hand, when operating on practical channels plagued with noise, a beam search decoder employing infinitely many beams is equivalent to Viterbi decoding\footnote{This perspective is analogous to stack decoding of convolutional codes \cite{jelinekFastSequentialDecoding1969} with a sufficiently large stack.}, under the assumption that marker symbols guarantee the synchronization of the decoded sequence with the syndrome tree. Consequently, adjusting the number of beams allows us to explore intermediate FER regimes. Indeed, we strive to show that the performance of trellis-based decoders can be approached by \syndec~despite the latter being linearly complex in the length of the payload sequences, since the nanopore sequencing channel offers a sufficient amount of soft information that can be intelligently leveraged, allowing us to correct errors in raw reads by adopting a faster sequential approach.

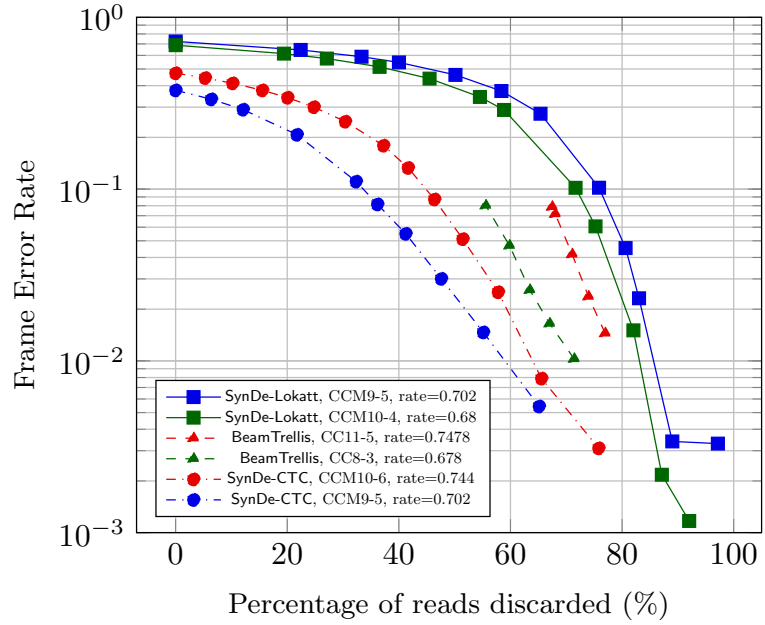
\begin{figure} 
    \centering
    \scalebox{1.2}{\begin{tikzpicture}
      \begin{axis}[legend style={nodes={scale=0.55, transform shape}},
				ymode=log,
				xmin=-7, xmax=105,
				ymin=1.001e-3, ymax=1.0,
				xlabel={\small Percentage of reads discarded (\%)}, ylabel={\small Frame Error Rate},
				legend pos=south west, legend style={font=\scriptsize},
				grid=both]
            \addplot[solid, blue!90!black, mark=square*, mark options={fill=blue!90!black}] table {resources/fer-discarded-reads-plots/volkel_dataset/FER_Lokatt_cc9_8_9_m5_r0.702.txt};
            \addlegendentry{\syndeclokatt, CCM9-5, rate=$0.702$};
            \addplot[solid, green!40!black, mark=square*, mark options={fill=green!40!black}] table {resources/fer-discarded-reads-plots/volkel_dataset/FER_Lokatt_cc11_10_10_m4_r0.68.txt};
            \addlegendentry{\syndeclokatt, CCM10-4, rate=$0.68$};
            \addplot[dashed, red!90!black, mark=triangle*, mark options={fill=red!90!black}] table {resources/fer-discarded-reads-plots/volkel_dataset/FER_Stan_cc6_5_11_r0.75.txt};
            \addlegendentry{\chandakdec, CC11-5, rate=$0.7478$};
            \addplot[dashed, green!40!black, mark=triangle*, mark options={fill=green!40!black}] table {resources/fer-discarded-reads-plots/volkel_dataset/FER_Stan_cc4_3_8_r0.68.txt};
            \addlegendentry{\chandakdec, CC8-3, rate=$0.678$};
            \addplot[dash dot, red!90!black, mark=*, mark options={fill=red!90!black}] table {resources/fer-discarded-reads-plots/volkel_dataset/FER_CTC_cc13_12_10_m6_r0.744.txt};
            \addlegendentry{\syndecctc, CCM10-6, rate=$0.744$};
            \addplot[dash dot, blue!90!black, mark=*, mark options={fill=blue!90!black}] table {resources/fer-discarded-reads-plots/volkel_dataset/FER_CTC_cc9_8_9_m5_r0.702.txt};
            \addlegendentry{\syndecctc, CCM9-5, rate=$0.702$};
        \end{axis}
  \end{tikzpicture}}
    \caption{Comparison of the decoding performance of \syndecctc~and \syndeclokatt~with \chandakdec~\cite{chandakOvercomingHighNanopore2020} on the dataset from Volkel \textit{et al.}.  Extended version of Fig.~\ref{fig:fer-vs-discarded-reads}.}\label{supp:fig:fer-vs-discarded-reads-volkel-data}
\end{figure}

Fig.~\ref{supp:fig:fer-vs-discarded-reads-volkel-data}, Fig.~\ref{supp:fig:fer-vs-discarded-reads-lokatt-data}, and Fig.~\ref{supp:fig:fer-vs-discarded-reads-stanford-data} present the results of decoding simulations on the datasets by Volkel \textit{et al.}, Xu \textit{et al.}, and  Lau \textit{et al.}, respectively. 
In all figures, the accuracies of \chandakdec~and \syndec~are evaluated by examining how their respective FERs vary with the fraction of decoder outputs that are discarded on the basis of likelihood scores. 
We observe in Fig.~\ref{supp:fig:fer-vs-discarded-reads-volkel-data} that on the dataset by Volkel \textit{et al.},\syndecctc~outperforms \chandakdec. 
This may be attributed to the use of marker codes, which greatly mitigate the possibility of insertion and deletion errors that are so prevalent in traditional basecalling. 
Very interestingly, Fig.~\ref{supp:fig:fer-vs-discarded-reads-volkel-data} also indicates that more redundancy is needed for synchronization than for substitution correction.

\syndeclokatt~approaches the performance of \chandakdec, but fails to outperform it, possibly owing to the inherent limitations of the Lokatt basecaller. However, it is worth noting that since both \syndecctc~and \chandakdec~heavily rely on the architecture of CTC basecallers, a comparison between them is fairer than one between \syndeclokatt~and \chandakdec. In fact, in our simulations, \syndecctc~and \chandakdec~receive soft information on the raw reads from the same neural network used by Bonito basecallers for flow cell chemistry R9.4.1.

In Fig.~\ref{supp:fig:fer-vs-discarded-reads-lokatt-data}, we observe that \syndeclokatt~is able to match the performance of \chandakdec~when both use codes of the same rate, while \syndecctc~outperforms both.

On the contrary, the results on the dataset by Lau \emph{et al.} (Fig.~\ref{supp:fig:fer-vs-discarded-reads-stanford-data}) indicate that \syndecctc~does not outperform \chandakdec~as it did on the other two datasets, but remains comparable despite being significantly more efficient.


\begin{figure} 
    \centering
    \scalebox{1.2}{\begin{tikzpicture}
      \begin{axis}[legend style={nodes={scale=0.52, transform shape}},
				ymode=log,
				xmin=-7, xmax=105,
				ymin=2.001e-4, ymax=1.0,
				xlabel={\small Percentage of reads discarded (\%)}, ylabel={\small Frame Error Rate},
				legend pos=south west, legend style={font=\scriptsize},
				grid=both]
            \addplot[dashed, red!90!black, mark=triangle*, mark options={fill=red!90!black}] table {resources/fer-discarded-reads-plots/lokatt_dataset/FER_Stan_cc6_5_11_r0.75.txt};
            \addlegendentry{\chandakdec, CC9-14, rate=$0.7478$};
            \addplot[solid, red!90!black, mark=square*, mark options={fill=red!90!black}] table {resources/fer-discarded-reads-plots/lokatt_dataset/FER_Lokatt_cc13_12_10_m6_r0.744.txt};
            \addlegendentry{\syndeclokatt, CCM10-6, rate=$0.744$};
            \addplot[dashed, blue!90!black, mark=triangle*, mark options={fill=blue!90!black}] table {resources/fer-discarded-reads-plots/lokatt_dataset/FER_Stan_cc4_3_8_r0.68.txt};
            \addlegendentry{\chandakdec, CC10-6, rate=$0.678$};
            \addplot[solid, blue!90!black, mark=square*, mark options={fill=blue!90!black}] table {resources/fer-discarded-reads-plots/lokatt_dataset/FER_Lokatt_cc11_10_10_m4_r0.684.txt};
            \addlegendentry{\syndeclokatt, CCM10-4, rate=$0.68$};
            \addplot[dash dot, red!90!black, mark=*, mark options={fill=red!90!black}] table {resources/fer-discarded-reads-plots/lokatt_dataset/FER_CTC_cc13_12_10_m6_r0.744.txt};
            \addlegendentry{\syndecctc, CCM10-6, rate=$0.744$};
            \addplot[dash dot, blue!90!black, mark=*, mark options={fill=blue!90!black}] table {resources/fer-discarded-reads-plots/lokatt_dataset/FER_CTC_cc11_10_10_m4_r0.684.txt};
            \addlegendentry{\syndecctc, CCM10-4, rate=$0.684$};
        \end{axis}
  \end{tikzpicture}}
    \caption{Comparison of the decoding performance of \syndecctc~and \syndeclokatt~with \chandakdec~\cite{chandakOvercomingHighNanopore2020} on the dataset from Xu \textit{et al.}, similar to Fig.~\ref{fig:fer-vs-discarded-reads}. The FER performance of \chandakdec~is matched by \syndeclokatt\ and surpassed by \syndecctc. 
    }\label{supp:fig:fer-vs-discarded-reads-lokatt-data}
\end{figure}

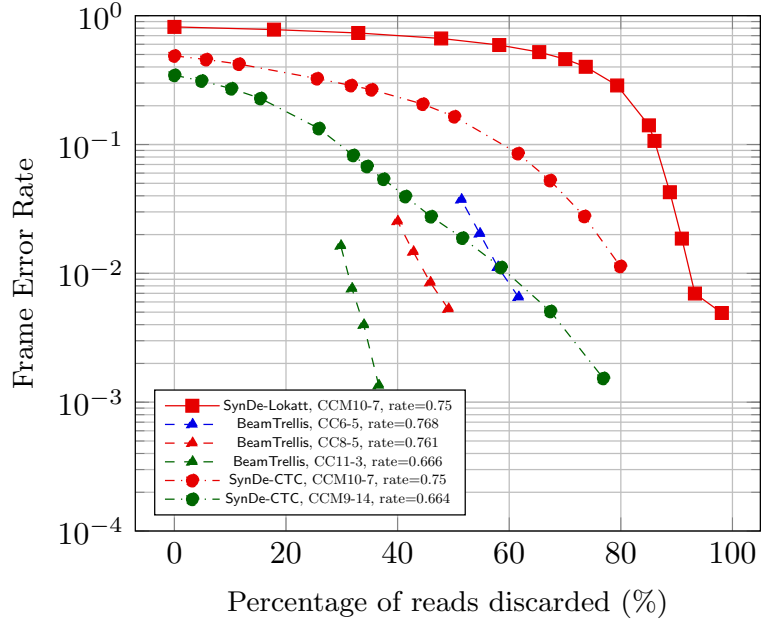
\begin{figure} 
    \centering
    \scalebox{1.2}{\begin{tikzpicture}
      \begin{axis}[legend style={nodes={scale=0.5, transform shape}},
				ymode=log,
				xmin=-7, xmax=105,
				ymin=1.001e-4, ymax=1.0,
				xlabel={\small Percentage of reads discarded (\%)}, ylabel={\small Frame Error Rate},
				legend pos=south west, legend style={font=\scriptsize},
				grid=both]
            \addplot[solid, red!90!black, mark=square*, mark options={fill=red!90!black}] table {resources/fer-discarded-reads-plots/stanford_dataset/FER_Lokatt_pslokatt_cc13_12_10_m7_r0.75.txt};
            \addlegendentry{\syndeclokatt, CCM10-7, rate=$0.75$};
            \addplot[dashed, blue!90!black, mark=triangle*, mark options={fill=blue!90!black}] table {resources/fer-discarded-reads-plots/stanford_dataset/FER_Stan_cc6_5_6_r0.768.txt};
            \addlegendentry{\chandakdec, CC6-5, rate=$0.768$};
            \addplot[dashed, red!90!black, mark=triangle*, mark options={fill=red!90!black}] table {resources/fer-discarded-reads-plots/stanford_dataset/FER_Stan_cc6_5_8_r0.761.txt};
            \addlegendentry{\chandakdec, CC8-5, rate=$0.761$};
            \addplot[dashed, green!40!black, mark=triangle*, mark options={fill=green!40!black}] table {resources/fer-discarded-reads-plots/stanford_dataset/FER_Stan_cc4_3_11_r0.666.txt};
            \addlegendentry{\chandakdec, CC11-3, rate=$0.666$};
            \addplot[dash dot, red!90!black, mark=*, mark options={fill=red!90!black}] table {resources/fer-discarded-reads-plots/stanford_dataset/FER_CTC_cc13_12_10_m7_r0.75.txt};
            \addlegendentry{\syndecctc, CCM10-7, rate=$0.75$};
            \addplot[dash dot, green!40!black, mark=*, mark options={fill=green!40!black}] table {resources/fer-discarded-reads-plots/stanford_dataset/FER_CTC_cc4_3_9_m14_r0.664.txt};
            \addlegendentry{\syndecctc, CCM9-14, rate=$0.664$};
        \end{axis}
  \end{tikzpicture}}
    \caption{Comparison of the decoding performance of \syndecctc~and \syndeclokatt~with \chandakdec~\cite{chandakOvercomingHighNanopore2020} on the dataset from Lau \textit{et al.}, similar to Fig.~\ref{fig:fer-vs-discarded-reads}. Unlike in Fig.~\ref{fig:fer-vs-discarded-reads} and Fig.~\ref{supp:fig:fer-vs-discarded-reads-lokatt-data}, the FER performance of \syndecctc~does not surpass that of \chandakdec, but remains comparable despite the lower complexity of \syndecctc.}\label{supp:fig:fer-vs-discarded-reads-stanford-data}
\end{figure}

\section{Complexity Analysis} \label{supp:complexity}

In the following, we present the time complexity of our algorithms \ps~and \syndec, both of which primarily involve many rounds of beam search operations. Recall from Fig.~\ref{fig:primer_search} and Fig.~\ref{fig:beam-search-decoder} that the process of beam search comprises $M$ rounds of beam extensions, along with merging, pruning, and sorting operations, where $M$ is proportional to the length of the raw read. Since the complexity of beam extensions comfortably dominate those of the sorting, merging, and pruning processes, we characterize the time complexity of both \ps~and \syndec~in terms of the average number of beam extensions performed for each sample of the raw signal. Henceforth, we refer to this as the \emph{mean beam complexity}.

\subsection{\ps} \label{supp:complexity-primer-search}

The notion of mean beam complexity is particularly useful for comparing with the primer localization approach adopted by Chandak \textit{et al.}, since the complexity of the latter is primarily influenced by that of basecalling, the mean beam complexity of which can be measured.

We conduct simulations on the dataset by Lau \textit{et al.} to measure the mean beam complexity of \psctc~as well as CTC basecalling. For the latter, we set $W=5$, which is the default number of beams used by the Bonito basecaller and in Chandak \textit{et al.}'s implementation, resulting in a mean beam complexity of $28.9$. On the other hand, running \psctc~with $W=8$, $s=6$, and $\tau=0.98$ yielded a mean beam complexity of $33.28$, such that for $82\%$ of the reads, the primer location estimates from both methods are within $50$ samples of each other. Notably, when using $W = 8$, $s = 1$, and $\tau = 1.0$, the two methods produced estimates within $50$ samples for $79.4\%$ of reads, but the mean beam complexity of \psctc~increased to $112.61$.

\subsection{\syndec} \label{supp:complexity-synde}


As we did for \ps, we now seek to demonstrate the computational advantage of \syndec~over \chandakdec. Although the latter does not involve beam search, it requires computing the probabilities of nodes over an expanded form of the convolutional code trellis \cite{chandakOvercomingHighNanopore2020}. These computations may be considered to be equivalent to the beam extensions used to define mean beam complexity. As a consequence, we adapt the definition of the mean beam complexity of \chandakdec~to be the average number of probability computations performed for each sample of the raw signal (or rather, the number of columns in the probability matrix $\boldsymbol{P}$). Essentially, this translates to the ratio of the number of edges in the expanded code trellis to the number of columns in $\boldsymbol{P}$. \syndecctc~is run with $W=512$ and the mean beam complexity figures of \syndecctc~and \chandakdec~for different coding schemes on the dataset by Lau \textit{et al.} are stated in Table~\ref{tab:codes1}
and Table~\ref{tab:codes2}, respectively. Therein, we observe that for codes with nearly identical rates, \syndecctc~offers a complexity reduction of at least two orders of magnitude compared to \chandakdec. 
For instance, when the error correction scheme CCM9-14 (Table~\ref{tab:codes1}) with rate $0.664$ and codeword length $113$ is used with \syndecctc, the mean beam complexity is $2.05 \times 10^3$, three orders of magnitude lower than the mean beam complexity of \chandakdec~when used to decode codewords from the code CC11-3 (Table~\ref{tab:codes2}) with rate $0.666$ and codeword length $117$.


To explicitly demonstrate how the time complexity of \syndec~and \chandakdec~are influenced by different parameter choices and convolutional codes, we also present the following theoretical analysis. First, we recall that \syndec~initiates a single beam from the raw sample from where the primer sequence is believed to have started, as depicted in Fig.~\ref{fig:beam-search-decoder}. Subsequently, this beam is propagated through a `primer region', then a `payload region', and finally another primer region. While the number of extension events that every beam undergoes, i.e., child beams created per parent beam in the two primer regions is usually $2$, corresponding to a dwelling event and an extension event,  respectively, the payload region involves the creation of $\rho+1$ child beams for each parent beam, where $\rho$ refers to the average number of outgoing edges per level in the quaternary syndrome trellis of the convolutional code being used. Assuming the computational cost of a creating single child beam to be $C_1$ and that at most $W$ beams are employed in each iteration, we infer that the net mean beam complexity of \syndec~can be expressed as follows.
\begin{align*}
    \text{Mean beam complexity of \syndeclokatt} &\propto 2\cdot 2C_1W dL + (\rho+1)C_1W dM \\
    &= (4L+(\rho+1)M)C_1Wd,
\end{align*}
where $d$ refers to the average dwell time of a single $k$-mer in the nanopore, while $L$ and $M$ refer to the lengths of the primer and the payload sequences, respectively. 

For CTC basecallers like Bonito which typically involve a stride parameter, this expression simply transforms into:
\begin{align*}
    \text{Mean beam complexity of \syndecctc} &\propto (4L+(\rho+1)M)C_1Wd/\sigma,
\end{align*}
since the number of columns in the associated neural network's output probability matrix $\boldsymbol{P}$ is at most $1/\sigma$ times the length of the raw signal.

Note that for any linear error correction code, it holds that $\rho < 4$, since $\rho=4$ implies that every parent beam may undergo extension to any of the $4$ bases of the DNA alphabet, as in basecalling. Thus, the structural complexity of the tree explored by \syndec's beam search is strictly less than that of the tree traversed by a basecaller's beam search. 



In contrast, the mean beam complexity of \chandakdec~\cite{volkel_nanopore_2025} can be characterized as:
\begin{align}
    \text{Complexity of \chandakdec} &\approx C_2M^{2}N2^\nu/\sigma, \label{eq::stanford-dec-complexity}
\end{align}
where $N$ refers to the average length of the raw signal that corresponds exclusively to the payload (of length $M$), $C_2$ refers to the cost of a single score computation, and $\nu$ represents the memory of the convolutional code. Understandably, $N$ is proportional to $M$, and the stride $\sigma$ appears due to the use of the Bonito basecaller. 

Table~\ref{supp:tab:complexity} compares the time and storage complexities of \syndec, \chandakdec, and another competing decoding algorithm called \volkeldec~\cite{volkel_nanopore_2025}. Although \volkeldec~remarkably improves on \chandakdec~in terms of efficiency, its time and storage complexities still scale exponentially with $\nu$. Thus it becomes evident that unlike \chandakdec~and \volkeldec, \syndec~is capable of accommodating codes of higher memory without incurring any significant increase in decoding latency, which we rely on in our performance comparisons in Section~\ref{sec:results}.


\begin{table}[]
    \centering
    \scalebox{0.85}{\begin{tabular}{c c c}
    \toprule
     & Time Complexity & Storage Complexity \\
    \midrule
    \syndec~  & $O(WN)$  & $O(WM)$ \\
    \chandakdec~\cite{chandakOvercomingHighNanopore2020, lauMagneticDNARandom2023} & $O(M^2N2^{\nu})$ & $O(M^22^{\nu})$ \\
    \volkeldec~\cite{volkel_nanopore_2025} & $O(MN2^{\nu})$ & $O(N2^\nu )$ \\
    \bottomrule
\end{tabular}}
    \caption{Comparison of time and storage complexities. Here $W$ denotes the number of beams and $\nu$ refers to the memory of the convolutional code being used. $L$, $M$, and $N$ denote the lengths of the primer sequence, the payload, and the raw read,  respectively. Understandably, $N$ is proportional to $M$. In this summary, we have ignored the complexity of sorting $W$ beams as it is comfortaby exceeded by that of the beam-hghgextension phase in beam search algorithms. 
    }
     \label{supp:tab:complexity}
\end{table}

\end{document}

%% file: resources/macros.tex
\newcommand{\bfc}{{\boldsymbol c}}

\newcommand{\bfh}{{\boldsymbol h}}

\newcommand{\bfzero}{{\mathbf 0}}

\newcommand{\cC}{\mathcal{C}}

\newcommand{\chandakdec}{\textsf{BeamTrellis}}
\newcommand{\volkeldec}{\textsf{AlignmentMatrix}}
\newcommand{\ps}{\textsf{PrimerSeeker}}
\newcommand{\psctc}{\textsf{PrimerSeeker-CTC}}
\newcommand{\pslokatt}{\textsf{PrimerSeeker-Lokatt}}
\newcommand{\syndec}{\textsf{SynDe}}
\newcommand{\syndeclokatt}{\textsf{SynDe-Lokatt}}
\newcommand{\syndecctc}{\textsf{SynDe-CTC}}



%% file: resources/decoding-pipeline/pipeline.tex
\tikzset{ 	
    my_block/.style={ 			
        draw=black,
        fill=green!15,
        thick,
        rounded corners,
        minimum width=3cm,
        minimum height=1.5cm,
        align=center 	
    } 	
}  

\newcommand{\drawblock}[3]{\node[my_block,name=#1] at #2 {#3};}

\begin{tikzpicture}[>=stealth, node distance=2cm]   
    
    \node (sig) [draw=none,name=block1] at (0,0) {\includegraphics[scale=0.3]{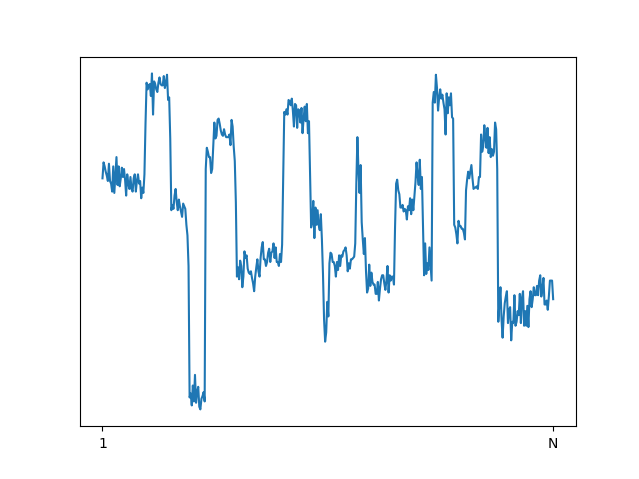}};
    \drawblock{block2}{(5cm,0)}{Neural Network}
    \draw[->, thick] (block1.east) -- (block2.west);
    
    \drawblock{block3}{(12cm,0)}{\ps}
    \draw[->, thick] (block2.east) -- (block3.west);
    
    \path (block2.east) -- (block3.west) coordinate[midway] (mid23arrow);
    \node (heat1) [above=0.3cm of mid23arrow] {
        \begin{adjustbox}{valign=t}
            \begin{tikzpicture}[scale=0.8]
                \foreach \i in {0,...,5} {
                    \foreach \j in {0,...,7} {
                        \pgfmathsetmacro{\randval}{int(mod(\i*13+\j*7+3,33))}
                        \global\expandafter\edef\csname randval\i\j\endcsname{\randval}
                        \pgfmathsetmacro{\colorpct}{\randval*100/32}
                        \pgfmathsetmacro{\colorval}{max(0,min(100,int(\colorpct)))}
                        \pgfmathparse{int(\colorval)}
                        \let\cval\pgfmathresult
                        \ifnum\cval<34
                        \pgfmathsetmacro{\mixpct}{max(0,min(100,\cval*3))}
                        \fill[black!\mixpct!blue] (\j*0.4,-\i*0.4) rectangle ++(0.4,-0.4);
                        \else
                        \ifnum\cval<67
                        \pgfmathsetmacro{\mixpct}{max(0,min(100,(\cval-34)*3))}
                        \fill[blue!\mixpct!green] (\j*0.4,-\i*0.4) rectangle ++(0.4,-0.4);
                        \else
                        \pgfmathsetmacro{\mixpct}{max(0,min(100,(\cval-67)*3))}
                        \fill[green!\mixpct!yellow] (\j*0.4,-\i*0.4) rectangle ++(0.4,-0.4);
                        \fi
                        \fi
                    }
                }
                \node at (0.2,0.1) {\tiny 1};
                \node at (3.0,0.1) {\tiny N};
                \node[left] at (-0.1,-0.2) {\tiny $S_1$};
                \node[left] at (-0.1,-2.2) {\tiny $S_M$};
                \node[left] at (-0.1,-0.6) {\tiny $S_2$};
                \node[left] at (-0.6,-1.25) {\tiny $\vdots$};
            \end{tikzpicture}
        \end{adjustbox}
    };

    \drawblock{block4}{(12cm,-3cm)}{Crop}
    
    \draw[->, thick, dashed] (block3.south) -- node[right] {$n_{\mathrm{P}}$} (block4.north);
    
    \path (block2.east) -- (block3.west) coordinate[midway] (mid23);
    \draw[->, thick] (mid23) |- (block4.west);

    \node (block5) [draw=none, align=center, above=1.5cm of block3] {Start Primer \\ $\dA\dC\dG\dT$..};
    \draw[->, thick] (block5.south) -- (block3.north);
    
    \drawblock{block6}{(12cm,-7.5cm)}{\syndec}
    
    \draw[->, thick] (block4.south) -- (block6.north);
    \path (block4.south) -- (block6.north) coordinate[midway] (mid46);
    \node (heat2) [left=0.3cm of mid46] {
        \begin{adjustbox}{valign=t}
            \begin{tikzpicture}[scale=0.8]
                \foreach \i in {0,...,5} {
                    \foreach \joriginal [count=\jnew from 0] in {3,4,5,6,7} {
                        \pgfmathsetmacro{\randval}{\csname randval\i\joriginal\endcsname}
                        \pgfmathsetmacro{\colorpct}{\randval*100/32}
                        \pgfmathsetmacro{\colorval}{max(0,min(100,int(\colorpct)))}
                        \pgfmathparse{int(\colorval)}
                        \let\cval\pgfmathresult
                        \ifnum\cval<34
                        \pgfmathsetmacro{\mixpct}{max(0,min(100,\cval*3))}
                        \fill[black!\mixpct!blue] (\jnew*0.4,-\i*0.4) rectangle ++(0.4,-0.4);
                        \else
                        \ifnum\cval<67
                        \pgfmathsetmacro{\mixpct}{max(0,min(100,(\cval-34)*3))}
                        \fill[blue!\mixpct!green] (\jnew*0.4,-\i*0.4) rectangle ++(0.4,-0.4);
                        \else
                        \pgfmathsetmacro{\mixpct}{max(0,min(100,(\cval-67)*3))}
                        \fill[green!\mixpct!yellow] (\jnew*0.4,-\i*0.4) rectangle ++(0.4,-0.4);
                        \fi
                        \fi
                    }
                }

                \node at (0.6,0.1) {\tiny $n_{\mathrm{P}}$};
                \node at (2.1,0.1) {\tiny N};

                \node[left] at (-0.1,-0.2) {\tiny $S_1$};
                \node[left] at (-0.1,-2.2) {\tiny $S_M$};
                \node[left] at (-0.1,-0.6) {\tiny $S_2$};
                \node[left] at (-0.6,-1.25) {\tiny $\vdots$};
            \end{tikzpicture}
        \end{adjustbox}
    };

    \node (block7) [draw=none, align=center, right=1cm of block6] {Primer Pair \\ ($\dA\dC\dG\dT$..., $\dT\dC\dG\dG$...)};
    \draw[->, thick] (block7.west) -- (block6.east);

    \node (block8) [draw=none, align=center, left=3cm of block6] {\textbf{Decoded Sequence}};
    \draw[->, thick] (block6.west) -- (block8.east);
    \node (decseq) [draw=none, below=0.1cm of block8] { {$\dC\dA\dC\dG\dT\dA$}..};
\end{tikzpicture}

%% file: resources/primer-search-pipeline/primer_search_alg.tex


\tikzset{
	child_beam/.style={
		draw=red!15!black,
		thick,
		ellipse,
		fill=red!15,
		minimum width=1.2cm,
		minimum height=0.8cm,
		align=center
	},
	parent_beam/.style={
		draw=blue!15!black,
		thick,
		ellipse,
		fill=blue!15,
		minimum width=1.2cm,
		minimum height=0.8cm,
		align=center
	},
	child_edge/.style={
		->,
		>=stealth,
		thick,
		black
	},
	carry_edge/.style={
		dashed,
		gray,
		thick
	}
}
\tikzset{ 	
    my_block/.style={ 			
        draw=black,
        fill=green!15,
        thick,
        rounded corners,
        minimum width=3cm,
        minimum height=1.5cm,
        align=center 	
    } 	
}  

\newcommand{\drawblock}[3]{\node[my_block,name=#1] at #2 {#3};}

\newcommand{\childbeam}[3]{\node[child_beam] (#1) at #2 {#3};}
\newcommand{\parentbeam}[3]{\node[parent_beam] (#1) at #2 {#3};}
\newcommand{\childedge}[4][]{\draw[child_edge] #2 -- #3 node[midway, above, font=\normalsize] {#4} #1;}
\newcommand{\childedgebelow}[4][]{\draw[child_edge] #2 -- #3 node[midway, below, font=\normalsize] {#4} #1;}
\newcommand{\childedgeboth}[5][]{\draw[child_edge] #2 -- #3 node[midway, above, sloped, font=\normalsize] {#4} node[midway, below, sloped, font=\normalsize] {#5} #1;}
\newcommand{\carryedge}[2]{\draw[carry_edge] #1 -- #2;}

	\begin{tikzpicture}[->,>=stealth,node distance=2.5cm and 2cm]
		\def\hsep{4.7cm}
		\def\hsepp{3.4cm}
		\def\vsep{9cm}
		\def\vsepp{1.3cm}
		
		\parentbeam{start_a}{(0,0)}{start = $a$}
		\parentbeam{start_b}{(0,-\vsep)}{start = $b$}
		
		
				
		\parentbeam{s1_a_ca_parent}{($(start_a) + (\hsepp,0)$)}{start = $a$\\$\dC\dA$}
		\parentbeam{s1_b_ca_parent}{($(start_b) + (\hsepp,0)$)}{start = $b$\\$\dC\dA$} 
		
		\carryedge{(start_a)}{(s1_a_ca_parent)}
		\carryedge{(start_b)}{(s1_b_ca_parent)}
		
		\childbeam{s1_a_ca_dwell}{($(s1_a_ca_parent) + (\hsep,2cm)$)}{start = $a$\\$\dC\dA$}
		\childbeam{s1_a_ca_ext}{($(s1_a_ca_parent) + (\hsep,-2cm)$)}{start = $a$\\$\dA\dC$}
		
		\childbeam{s1_b_ca_dwell}{($(s1_b_ca_parent) + (\hsep,2cm)$)}{start = $b$\\$\dC\dA$}
		\childbeam{s1_b_ca_ext}{($(s1_b_ca_parent) + (\hsep,-2cm)$)}{start = $b$\\$\dA\dC$}
		
		\childedgeboth{(s1_a_ca_parent)}{(s1_a_ca_dwell)}{dwell at}{sample ${x_{a+1}}$}
		\childedgeboth{(s1_a_ca_parent)}{(s1_a_ca_ext)}{extend at}{sample ${x_{a+1}}$}
		
		\childedgeboth{(s1_b_ca_parent)}{(s1_b_ca_dwell)}{dwell at}{sample ${x_{b+1}}$}
		\childedgeboth{(s1_b_ca_parent)}{(s1_b_ca_ext)}{extend at}{sample ${x_{b+1}}$}
		
		\parentbeam{s2_a_ca}{($(s1_a_ca_dwell) + (\hsepp,0)$)}{start = $a$\\$\dC\dA$}
		\parentbeam{s2_a_ac}{($(s1_a_ca_ext) + (\hsepp,0)$)}{start = $a$\\$\dA\dC$} 
		\parentbeam{s2_b_ca}{($(s1_b_ca_dwell) + (\hsepp,0)$)}{start = $b$\\$\dC\dA$}
		\parentbeam{s2_b_ac}{($(s1_b_ca_ext) + (\hsepp,0)$)}{start = $b$\\$\dA\dC$}

		\carryedge{(s1_a_ca_dwell)}{(s2_a_ca)}
		\carryedge{(s1_a_ca_ext) }{(s2_a_ac)}
		\carryedge{(s1_b_ca_dwell)}{(s2_b_ca)}
		\carryedge{(s1_b_ca_ext)}{(s2_b_ac)}
		
		\childbeam{s2_a_ca_dwell}{($(s2_a_ca) + (\hsep,\vsepp)$)}{start = $a$\\$\dC\dA$}
		\childbeam{s2_a_ca_ext}{($(s2_a_ca) + (\hsep,-\vsepp)$)}{start = $a$\\$\dA\dC$}
		\childedgeboth{(s2_a_ca)}{(s2_a_ca_dwell)}{dwell at}{sample ${x_{a+2}}$}
		\childedgeboth{(s2_a_ca)}{(s2_a_ca_ext)}{extend at}{sample ${x_{a+2}}$}
		
		\childbeam{s2_a_ac_dwell}{($(s2_a_ac) + (\hsep,\vsepp)$)}{start = $a$\\$\dA\dC$}
		\childbeam{s2_a_ac_ext}{($(s2_a_ac) + (\hsep,-\vsepp)$)}{start = $a$\\$\dC\dG$}
		\childedgeboth{(s2_a_ac)}{(s2_a_ac_dwell)}{dwell at}{sample ${x_{a+2}}$}
		\childedgeboth{(s2_a_ac)}{(s2_a_ac_ext)}{extend at}{sample ${x_{a+2}}$}
		
		\childbeam{s2_b_ca_dwell}{($(s2_b_ca) + (\hsep,\vsepp)$)}{start = $b$\\$\dC\dA$}
		\childbeam{s2_b_ca_ext}{($(s2_b_ca) + (\hsep,-\vsepp)$)}{start = $b$\\$\dA\dC$}
		\childedgeboth{(s2_b_ca)}{(s2_b_ca_dwell)}{dwell at}{sample ${x_{b+2}}$}
		\childedgeboth{(s2_b_ca)}{(s2_b_ca_ext)}{extend at}{sample ${x_{b+2}}$}
		
		\childbeam{s2_b_ac_dwell}{($(s2_b_ac) + (\hsep,\vsepp)$)}{start = $b$\\$\dA\dC$}
		\childbeam{s2_b_ac_ext}{($(s2_b_ac) + (\hsep,-\vsepp)$)}{start = $b$\\$\dC\dG$}
		\childedgeboth{(s2_b_ac)}{(s2_b_ac_dwell)}{dwell at}{sample ${x_{b+2}}$}
		\childedgeboth{(s2_b_ac)}{(s2_b_ac_ext)}{extend at}{sample ${x_{b+2}}$}
		
		\parentbeam{s3_a_ca}{($(s2_a_ca_dwell) + (\hsep,0)$)}{start = $a$\\$\dC\dA$}
		\parentbeam{s3_a_cg}{($(s2_a_ac_ext) + (\hsep,0)$)}{start = $a$\\$\dC\dG$}
		\carryedge{(s2_a_ca_dwell)}{(s3_a_ca)}
		\carryedge{(s2_a_ac_ext) }{(s3_a_cg)}
		
		\parentbeam{s3_b_ca}{($(s2_b_ca_dwell) + (\hsep,0)$)}{start = $b$\\$\dC\dA$}
		\parentbeam{s3_b_cg}{($(s2_b_ac_ext) + (\hsep,0)$)}{start = $b$\\$\dC\dG$}
		\carryedge{(s2_b_ca_dwell)}{(s3_b_ca)}
		\carryedge{(s2_b_ac_ext) }{(s3_b_cg)}
		
		\parentbeam{s3_a_ac}{($(s2_a_ac_ext) + (\hsep,3.25cm)$)}{start = $a$\\$\dA\dC$}
		\parentbeam{s3_b_ac}{($(s2_b_ac_ext) + (\hsep,3.25cm)$)}{start = $b$\\$\dA\dC$}
		\carryedge{(s2_a_ca_ext)}{(s3_a_ac)}
		\carryedge{(s2_a_ac_dwell)}{(s3_a_ac)}
		\carryedge{(s2_b_ca_ext)}{(s3_b_ac)}
		\carryedge{(s2_b_ac_dwell)}{(s3_b_ac)}
		
		\foreach \cn in {s3_a_ca, s3_a_ac, s3_a_cg} {
			\childedgeboth{(\cn)}{($(\cn) + (2.0cm,0.8cm)$)}{}{}
			\childedgeboth{(\cn)}{($(\cn) + (2.0cm,-0.8cm)$)}{}{}
		}
		\foreach \cn in {s3_b_ca, s3_b_ac, s3_b_cg} {
			\childedgeboth{(\cn)}{($(\cn) + (2.0cm,0.8cm)$)}{}{}
			\childedgeboth{(\cn)}{($(\cn) + (2.0cm,-0.8cm)$)}{}{}
		}

		\node[draw=none, above=0.6cm of s3_a_ca,,font=\normalsize\bfseries] (level_3_current) {current beams};
		\node[draw=none,,font=\normalsize\bfseries] (level_2_next) at ($(level_3_current) + (-\hsep,0)$) {extended beams};
		\node[draw=none,,font=\normalsize\bfseries] (level_2_current) at ($(level_2_next) + (-\hsep,0)$) {current beams};
		\node[draw=none,font=\normalsize\bfseries] (level_1_next) at ($(level_2_current) + (-\hsepp,0)$) {extended beams};
		\node[draw=none,align=center,,font=\normalsize\bfseries] (level_1_current) at ($(level_1_next) + (-\hsep,0)$) {extend by \\ first $k$-mer\\at sample $x_{\text{start}}$};
		\node[draw=none,align=center,name=initbeams,font=\normalsize\bfseries] (level_0_current) at ($(level_1_current) + (-\hsepp,0)$) {initial empty \\ beams};

		\node[draw=none, align=center,xshift=0.17cm,yshift=-1.0cm] at ($($(s2_a_ca_ext)!0.5!(s2_a_ac_dwell)$)!0.5!(s3_a_ac)$) {\normalsize{\bfseries merge}\\\normalsize{\bfseries identical beams}};
				
        \node[draw=none, align=center, xshift=0.17cm,yshift=-1.0cm] at ($($(s2_b_ca_ext)!0.5!(s2_b_ac_dwell)$)!0.5!(s3_b_ac)$) {\normalsize{\bfseries merge}\\\normalsize{\bfseries identical beams}};


        \node[draw=none,align=center,name=rawsig,yshift=4cm,xshift=-1cm] (rawsig) at (level_1_current) {\includegraphics{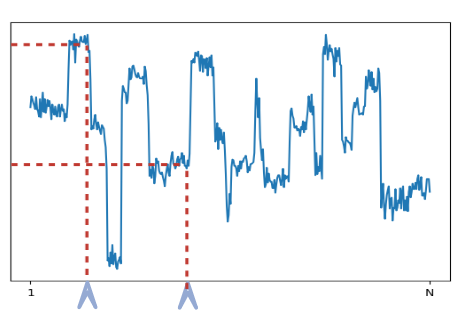}};
        \node[draw=none, scale=1.0, yshift=-1.5cm, xshift=-1.2cm] (ta) at (rawsig) {\scriptsize $t=a$};
        \node[draw=none, scale=1.0, xshift=-3mm, yshift=-1.5cm] (tb) at (rawsig) {\scriptsize $t=b$};
        \node[draw=none, scale=1.0, yshift=0.9cm, xshift=-2.1cm] (xa) at (rawsig) {\scriptsize $x_a$};
        \node[draw=none, scale=1.0, yshift=-1mm, xshift=-2.1cm] (xb) at (rawsig) {\scriptsize $x_b$};
        
        \drawblock{nn}{($(rawsig) + (6cm,0)$)}{Neural Network}
        \draw[->, thick] (rawsig.east) -- (nn.west);

        \drawblock{primersearch}{($(nn) + (8cm,0)$)}{\ps}
        \draw[->, thick] (nn.east) -- (primersearch.west);
        \node[xshift=4cm, name=primerIndex] at (primersearch) {\large $n_{\mathrm{P}}$};
        \draw[->, thick] (primersearch.east) -- (primerIndex.west);

        \node[name=targetseqk,yshift=4cm] at (primersearch) {Target $k$-mer sequence: $\dC\dA,\dA\dC,\dC\dG,\dG\dT,\dG\dG$};
        \draw[->, thick]  (targetseqk.south) -- (primersearch.north);

        \node[name=targetseq,xshift=-8cm] at (targetseqk) {Target primer: $\dC\dA\dC\dG\dT\dA\dG\dG$};
        \draw[->, thick]  (targetseq.east) -- (targetseqk.west);
        
        \path (nn.east) -- (primersearch.west) coordinate[midway] (midc);
        \node (heat1) [above=0.3cm of midc] {
        \begin{adjustbox}{valign=t}
            \begin{tikzpicture}[scale=0.8]
                \foreach \i in {0,...,5} {
                    \foreach \j in {0,...,7} {
                        \pgfmathsetmacro{\randval}{int(mod(\i*13+\j*7+3,33))}
                        \global\expandafter\edef\csname randval\i\j\endcsname{\randval}
                        \pgfmathsetmacro{\colorpct}{\randval*100/32}
                        \pgfmathsetmacro{\colorval}{max(0,min(100,int(\colorpct)))}
                        \pgfmathparse{int(\colorval)}
                        \let\cval\pgfmathresult
                        \ifnum\cval<34
                        \pgfmathsetmacro{\mixpct}{max(0,min(100,\cval*3))}
                        \fill[black!\mixpct!blue] (\j*0.4,-\i*0.4) rectangle ++(0.4,-0.4);
                        \else
                        \ifnum\cval<67
                        \pgfmathsetmacro{\mixpct}{max(0,min(100,(\cval-34)*3))}
                        \fill[blue!\mixpct!green] (\j*0.4,-\i*0.4) rectangle ++(0.4,-0.4);
                        \else
                        \pgfmathsetmacro{\mixpct}{max(0,min(100,(\cval-67)*3))}
                        \fill[green!\mixpct!yellow] (\j*0.4,-\i*0.4) rectangle ++(0.4,-0.4);
                        \fi
                        \fi
                    }
                }
                \node at (0.2,0.1) {\tiny 1};
                \node at (3.0,0.1) {\tiny N};
                \node[left] at (-0.1,-0.2) {\tiny \dA\dA};
                \node[left] at (-0.1,-2.2) {\tiny \dT\dT};
                \node[left] at (-0.1,-0.6) {\tiny \dA\dC};
                \node[left] at (-0.6,-1.25) {\tiny $\vdots$};
            \end{tikzpicture}
        \end{adjustbox}
    };
    
    \node[draw=none,name=westp,yshift=6.5cm,xshift=-5cm] at (s1_a_ca_parent) {};
    \draw[-, thick] (westp) -- ($(westp)+(25cm,0)$);
	\end{tikzpicture}


%% file: resources/decoding-pipeline/syndrome-trellis.tex
\begin{tikzpicture}[long dash/.style={dash pattern=on 10pt off 2pt}, short dash/.style={ dash pattern=on 3pt off 2.4pt}, ->,>=stealth]

    \tikzstyle{tiny dot}=[circle, fill, inner sep=0.5pt]
    \tikzstyle{syndrome node}=[circle, inner sep=4pt, fill=brown!15,font=\small]
    \newlength{\vsep}
    \newlength{\hsep}
    \setlength{\vsep}{1.3cm}
    \setlength{\hsep}{1.5cm}



    \foreach \j/\i/\synstate in {0/0/0, 
                                1/0/0, 1/1/160,
                                2/0/0, 2/1/80,
                                3/0/0, 3/1/80, 3/2/100, 3/3/52,
                                4/0/0, 4/1/16, 4/2/36, 4/3/52,
                                5/0/0, 5/1/56, 5/2/36, 5/3/28,
                                6/0/0, 6/1/24, 6/2/4, 6/3/28,
                                7/0/0, 7/1/24, 7/2/29, 7/3/5,
                                8/0/0, 8/1/20,
                                9/0/0, 9/1/60,
                                10/0/0}{
            \node[draw, syndrome node] (x\j\synstate) at ($(\j*\hsep, -\i*\vsep)$){\synstate};
    }
    
    \foreach \ia/\syna/\ib/\synb in {2/0/3/0,3/0/4/0,4/0/5/0,5/0/6/0,6/0/7/0,7/0/8/0,
    2/80/3/80,3/52/4/52,4/36/5/36,5/28/6/28,6/24/7/24}{
        \draw[->, short dash] (x\ia\syna) -- (x\ib\synb);
    }      
    \draw[->, solid] (x00.south) to [out=-90,in=180] (x1160.west);
    \draw[->, solid] (x280.south) to [out=-90,in=180] (x352.west);    
    \draw[->, solid] (x380.east) to [out=0,in=180] (x436.west);
    \draw[->, solid] (x536.east) to [out=-5,in=180] (x624.west);
    \draw[->, short dash] (x75.east) to [out=5,in=265] (x820.south);
    \draw[->, solid] (x960.east) to [out=5,in=265] (x100.south);
    
    \foreach \ia/\syna/\ib/\synb in {1/160/2/80,4/16/5/56,4/52/5/28,6/28/7/5,7/24/8/20,8/20/9/60}{
        \draw[->, solid] (x\ia\syna) -- (x\ib\synb);
    }

    \draw[->, short dash, green!40!black] (x00) -- (x10);
    \draw[->, short dash, green!40!black] (x10) -- (x20);
    \draw[->, solid, green!40!black] (x20.east) to [out=-5,in=180] (x3100.west);
    \draw[->, solid, green!40!black] (x3100.east) to [out=0,in=180] (x416.west);
    \draw[->, solid, green!40!black] (x416) -- (x556);
    \draw[->, solid, green!40!black] (x556.east) to [out=-5,in=180] (x64.west);
    \draw[->, solid, green!40!black] (x64) -- (x729);
    \draw[->, solid, green!40!black] (x729.east) to [out=5,in=180] (x80.west);
    \draw[->, short dash, green!40!black] (x80) -- (x90);
    \draw[->, short dash, green!40!black] (x90) -- (x100);
    

\end{tikzpicture}

%% file: resources/decoding-pipeline/beam-search-decoder.tex


\tikzset{
	blue_beam/.style={
		draw=blue!15!black,
		thick,
		ellipse,
		fill=blue!15,
		minimum width=1.8cm,
		minimum height=1.8cm,
		align=center
	},
    red_beam/.style={
		draw=red!15!black,
		thick,
		ellipse,
		fill=red!15,
		minimum width=1.8cm,
		minimum height=1.8cm,
		align=center
	},
    gray_beam/.style={
		draw=gray!15!black,
		thick,
		ellipse,
		fill=gray!15,
		minimum width=1.8cm,
		minimum height=1.8cm,
		align=center
	},
}
\tikzset{ 	
    merge_block/.style={ 			
        draw=black,
        fill=green!15,
        thick,
        rounded corners,
        minimum width=3cm,
        minimum height=12cm,
        align=center 	
    } 	
}  
\newcommand{\mergeblock}[3]{\node[merge_block,name=#1] at #2 {#3};}
\newcommand{\bluebeam}[3]{\node[blue_beam] (#1) at #2 {#3};}
\newcommand{\redbeam}[3]{\node[red_beam] (#1) at #2 {#3};}
\newcommand{\graybeam}[3]{\node[gray_beam] (#1) at #2 {#3};}

\scalebox{0.5}{
\begin{tikzpicture}[->,>=stealth,node distance=2.5cm and 2cm]
    \def\hsep{4.7cm}
    \def\hsepp{3.4cm}
    \def\vsep{2.3cm}
    \def\vsepp{3cm}
    
    \bluebeam{start}{(0,0)}{$\dC\dA$}     
    \bluebeam{ca1}{($(start) + (\hsepp,\vsepp)$)}{$\dC\dA$}
    \bluebeam{at1}{($(start) + (\hsepp,-\vsepp)$)}{$\dA\dT$\\$S_0=0$}

    \graybeam{ca2}{($(ca1) + (\hsepp,\vsep)+(0,-0.95cm)$)}{$\dC\dA$}
    \graybeam{at2}{($(ca1) + (\hsepp,-\vsep)+(0,0.95cm)$)}{$\dA\dT$\\$S_0=0$}
    \graybeam{at3}{($(at1) + (\hsepp,\vsep)$)}{$\dA\dT$\\$S_0=0$}
    \graybeam{ta1}{($(at1) + (\hsepp,0)$)}{$\dT\dA$\\$S_1=0$}
    \graybeam{tt1}{($(at1) + (\hsepp,-\vsep)$)}{$\dT\dT$\\$S_1=20$}
    \mergeblock{merge1}{($(start)+(2.9*\hsepp,0)+(0,-0.5cm)$)}{Merge identical\\beams and\\keep $W=2$\\top-scoring ones}
    \bluebeam{at4}{($(ca1) + (2.9*\hsepp,-0.2cm)$)}{$\dA\dT$\\$S_0=0$}
    \bluebeam{tt4}{($(at1) + (2.9*\hsepp,-0.65cm)$)}{$\dT\dT$\\$S_1=20$}
    \graybeam{at5}{($(at4) + (\hsepp,\vsep)+(0,-0.3cm)$)}{$\dA\dT$\\$S_0=0$}
    \graybeam{ta4}{($(at4) + (\hsepp,0)$)}{$\dT\dA$\\$S_1=0$}
    \graybeam{tt5}{($(at4) + (\hsepp,-\vsep)+(0,0.3cm)$)}{$\dT\dT$\\$S_1=20$}
    \graybeam{tt6}{($(tt4) + (\hsepp,\vsep)+(0,-0.3cm)$)}{$\dT\dT$\\$S_1=20$}
    \graybeam{tc1}{($(tt4) + (\hsepp,0)$)}{$\dT\dC$\\$S_2=36$}
    \graybeam{tg1}{($(tt4) + (\hsepp,-\vsep)+(0,0.3cm)$)}{$\dT\dG$\\$S_2=52$}
    \mergeblock{merge2}{($(merge1)+(2.9*\hsepp,0)$)}{Merge identical\\beams and\\keep $W=2$\\top-scoring ones}
    \bluebeam{tt7}{($(ta4) + (1.9*\hsepp,0)$)}{$\dT\dT$\\$S_1=20$}
    \bluebeam{tc2}{($(tc1) + (1.9*\hsepp,0)$)}{$\dT\dC$\\$S_2=36$}

    \draw[->, dashed, red!60, thick] (start) -- (ca1);
    \draw[->, solid, red!60, thick] (start) -- (at1);
    \draw[->, dashed, red!60, thick] (ca1) -- (ca2);
    \draw[->, solid, red!60, thick] (ca1) -- (at2);
    \draw[->, dashed, black, thick] (at1) -- (at3);
    \draw[->, solid, black, thick] (at1) -- (ta1);
    \draw[->, solid, black, thick] (at1) -- (tt1);
    \foreach \startnode in {ca2,at2,at3,ta1,tt1}{
    \draw[->, dotted, black, thick] 
        (\startnode) -- ($(\startnode)+(1.6cm,0)$);
    }
    \foreach \endnode in {at4,tt4}{
        \draw[->, dotted, black, thick] ($(-1.9cm,0)+(\endnode)$) -- (\endnode);
    }  
    \draw[->, dashed, black, thick] (at4) -- (at5);
    \draw[->, solid, black, thick] (at4) -- (ta4);
    \draw[->, solid, black, thick] (at4) -- (tt5);
    \draw[->, dashed, black, thick] (tt4) -- (tt6);
    \draw[->, solid, black, thick] (tt4) -- (tc1);
    \draw[->, solid, black, thick] (tt4) -- (tg1);
    \foreach \startnode in {at5,ta4,tt5,tt6,tc1,tg1}{
        \draw[->, dotted, black, thick] 
        (\startnode) -- ($(\startnode)+(1.6cm,0)$);
    }  
    \foreach \endnode in {tt7,tc2}{
        \draw[->, dotted, black, thick] ($(-1.9cm,0)+(\endnode)$) -- (\endnode);
    }  
    \foreach \startnode in {tt7,tc2}{
        \draw[->, dotted, black, thick] (\startnode) -- ($(\startnode)+(1.5cm,0.9cm)$);
        \draw[->, solid, black, thick] (\startnode) -- ($(\startnode)+(1.5cm,0cm)$);
        \draw[->, solid, black, thick] (\startnode) -- ($(\startnode)+(1.5cm,-0.9cm)$);
    }  

    \node[align=center](header1) at ($(start)+(1.15cm,3*\vsep)$) {propagate beams\\through sample $x_n$};
    \node[align=center](header2) at ($(header1)+(\hsepp,0)+(0.2cm,0)$) {propagate beams\\through sample $x_{n+1}$};
    \node[align=center](header3) at ($(header2)+(3*\hsepp,0)$) {propagate beams\\through sample $x_{n+2}$};
\end{tikzpicture}
}